\documentclass[aps,prd,11pt]{revtex4}
\usepackage[colorlinks=true, pdfstartview=FitV, linkcolor=blue, citecolor=red, urlcolor=magenta]{hyperref}

\usepackage{graphicx}
\usepackage{latexsym}
\usepackage{amsmath}
\usepackage{amsfonts}
\usepackage{amssymb}
\usepackage{verbatim}
\usepackage{soul}

\usepackage{tikz,tikz-3dplot}
\tdplotsetmaincoords{80}{45}

\tikzset{surface/.style={draw=black, fill=white, fill opacity=.6}}

\usepackage{tikz}
\usetikzlibrary{decorations.pathmorphing,patterns}

\newcommand{\be}{\begin{equation}}
\newcommand{\ee}{\end{equation}}
\newcommand{\bea}{\begin{eqnarray}}
\newcommand{\eea}{\end{eqnarray}}

\newcommand{\ben}{\begin{eqnarray}}
\newcommand{\een}{\end{eqnarray}}

\begin{document}

\title{AdS/BCFT correspondence and Horndeski gravity in the presence of gauge fields: holographic paramagnetism/ferromagnetism phase transition}

\author{Fabiano F. Santos$^{a}$}
\email{fabiano.ffs23-at-gmail.com} 

\author{Mois\'es Bravo-Gaete$^{b}$}
\email{mbravo-at-ucm.cl} 

\author{Oleksii Sokoliuk $^{c,d}$}
\email{oleksii.sokoliuk-at-mao.kiev.ua}

\author{Alexander Baransky $^{c}$}
\email{abaransky-at-ukr.net}

\affiliation{$^{a}$Instituto de Física, Universidade Federal do Rio de Janeiro, Caixa Postal 68528, Rio de Janeiro-RJ, 21941-972 -- Brazil.}

\affiliation{$^{b}$ Departamento de Matem\'aticas, F\'isica y Estad\'istica. Facultad de Ciencias
B\'asicas, Universidad Cat\'olica del Maule, Casilla 617, Talca,
Chile.}

\affiliation{$^{c}$ Astronomical Observatory, Taras Shevchenko National University of Kyiv, 3 Observatorna St., 04053 Kyiv, Ukraine,}

\affiliation{$^{d}$Main Astronomical Observatory of the NAS of Ukraine (MAO NASU), Kyiv, 03143, Ukraine.}

\begin{abstract}
This paper presents a dual gravity model for a (2+1)-dimensional system with a limit on finite charge density and temperature, which will be used to {study} the properties of the holographic phase transition to paramagnetism-ferromagnetism in the presence of Horndeski gravity terms. In our model, the non-zero charge density is supported by a magnetic field. As a result, the radius $\rho/B$ indicates a localized condensate, as we increase the Horndeski gravity parameter, {that is represented by} $\gamma$. Furthermore, {such condensate} shows quantum Hall-type behavior. This radius is also inversely related to the total action coefficients of our model. It was observed that increasing the Horndeski parameter decreases the critical temperature of the holographic model and leads to the harder formation of the magnetic moment at the bottom of the black hole. However, when removing the magnetic field, the ferromagnetic material presents a disorder of its magnetic moments, which is observed through the entropy of the system. We also found that at low temperatures, spontaneous magnetization and ferromagnetic phase transition.
\end{abstract}

\maketitle


\section{Introduction}

\begin{figure}[!ht]
\begin{center}
\includegraphics[scale=0.22]{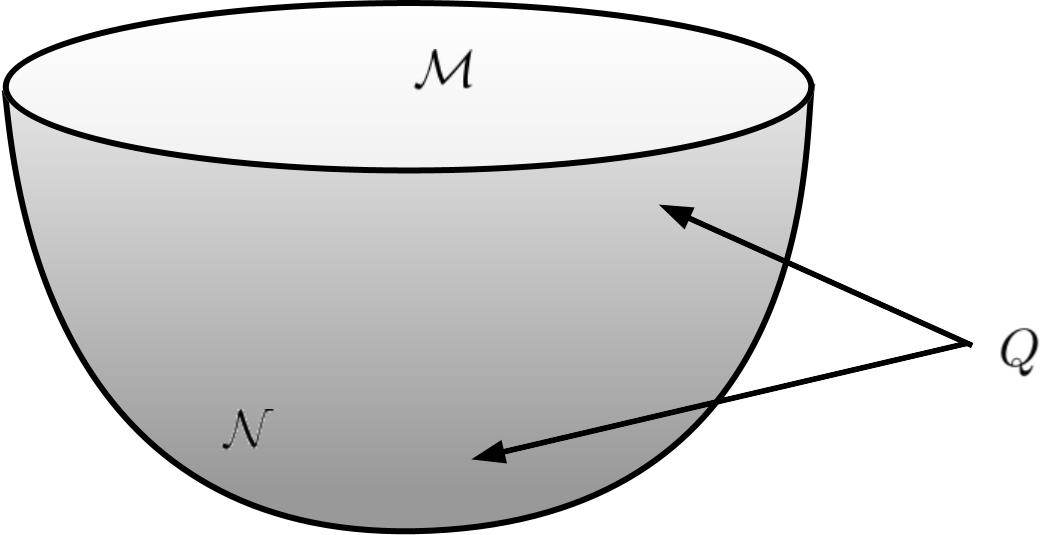}
\caption{Schematic representation of the AdS/BCFT correspondence. Here, $\mathcal{M}$ represents the manifold with boundary $\partial \mathcal{M}$ where the CFT is present. On the other hand, the gravity side is represented by $\mathcal{N}$, which is asymptotically AdS is $\mathcal{M}$. Together with the above, $\partial \mathcal{M}$ is extended into the bulk AdS, which constitutes the boundary of the $D-$dimensional manifold $Q$.}
\label{p}
\end{center}
\end{figure}

{For almost thirty years, the Anti-de Sitter/Conformal Field Theory (AdS/CFT) correspondence has been a bridge that allows us to relate gravity and strongly coupled conformal field theories \cite{Maldacena:1997re,Witten:1998qj}. Following this spirit, a new holographic dual of a CFT arises, which is defined on a manifold $\mathcal{M}$ with a boundary $\partial \mathcal{M}$, denoted as Boundary Conformal Field Theory (BCFT), proposed by Takayanagi \cite{Takayanagi:2011zk} and Takayanagi et al. \cite{Fujita:2011fp}, extending the AdS/CFT duality. This new holographic dual denoted as AdS/BCFT correspondence, is defined on a manifold
boundary in a $D$-dimensional manifold $\mathcal{M}$ to a $(D+1)$-dimensional asymptotically AdS space $\mathcal{N}$ in order to $\partial \mathcal{N}=\mathcal{M} \cup Q$. Here, $Q$ corresponds to a $D$-dimensional manifold that satisfies $\partial Q=\partial \mathcal{M}$ (see Figure \ref{p}).}

{At the moment to explore the AdS/CFT correspondence, we impose the Dirichlet boundary condition at the boundary of AdS,
and therefore we require the Dirichlet boundary condition on $\mathcal{M}$. Nevertheless, according to \cite{Takayanagi:2011zk,Fujita:2011fp}, for AdS/BCFT duality a Neumann
boundary condition (NBC) on $Q$ is required, given that this boundary should be
dynamical, from the viewpoint of holography, and there is no natural definite metric on $Q$ specified from the CFT side \cite{Nozaki:2012qd}.}

{On the other hand, the AdS/BCFT conjecture appears in many scenarios of the transport coefficients, where black holes take a providential role, such for example Hawking-Page phase transition, the Hall conductivity and the fluid/gravity correspondence  \cite{Fujita:2011fp,Melnikov:2012tb,dosSantos:2022scy,Miao:2018qkc,Sokoliuk:2022llp,Santos:2021orr,Magan:2014dwa}. Together with the above, this duality finds its natural roots in the holographic derivation of entanglement entropy \cite{Ryu:2006bv} as well as in the Randall-Sundrum model \cite{Randall:1999vf}. In fact, this extension of the CFT’s boundary inside the bulk of the AdS-space is considered a modification of a {\em thin} Randall-Sundrum brane, which intersects the AdS boundary. For this brane to be a dynamical object, we need to impose, as was shown before, NBC where the discontinuity in the bulk extrinsic curvature across the {\em defect}, is compensated by the tension from the brane. Furthermore, these boundaries are known as the Randall-Sundrum (RS) branes in the literature.}

{Following the above, Fujita et al. \cite{Fujita:2012fp} propose a model with gauge fields in the AdS$_{4}$ background with boundary RS branes. In this setup, the authors show that the additional boundary conditions impose relevant constraints on the gauge field parameters, deriving the Hall conductivity behavior in the dual field theory. Nevertheless, this approach does not consider the back reaction of the gauge fields on the geometry, constraining the geometry of the empty AdS space. A natural extension and generalization from the above work was constructed in \cite{Melnikov:2012tb}.}


{In the present paper, we are interested in constructing configurations  describing a physical system at finite temperature and charge density. For this, we consider the most common playground,  provided by the charged AdS$_{4}$ black holes. This background has} already been shown to encode many interesting condensed-matter-like phenomena such as superconductivity/superfluidity \cite{Gubser:2008px,Hartnoll:2008vx} and strange metallic behaviors \cite{Liu:2009dm}, {via an action characterized by the well-known Einstein-Hilbert structure together with a cosmological constant and Abelian gauge fields. It is interesting to note that the above toy model can be extended in the presence of boundaries within a special case of the Horndeski gravity \cite{Horndeski:1974wa}, (see for example \cite{Brito:2018pwe,Brito:2019ose,Santos:2019ljs,Santos:2020xox,Santos:2021guj,DosSantos:2022exb,Santos:2022lxj,Santos:2022uxm}). Here, the gravity theory is given through the Lagrangian 
\begin{eqnarray}\label{eq:Lhorn}
{\cal L}_{\rm H}= \kappa \Big[ (R-2\Lambda)\label{L1} -\frac{1}{2}(\alpha g_{\mu\nu}-\gamma\,  G_{\mu\nu})\nabla^{\mu}\phi\nabla^{\nu}\phi \Big],
\end{eqnarray}
where $R$, $G_{\mu \nu}$ and
$\Lambda$ are the scalar curvature, the Einstein tensor, and the cosmological constant respectively, $\phi=\phi(r)$ is a scalar field, $\alpha$ and $\gamma$ are coupling constants, while that $\kappa={1}/{(16 \pi G_N)}$, where $G_N$ is the Newton Gravitational constant. The Lagrangian (\ref{eq:Lhorn}) has been exhaustively explored from the perspective of hairy black hole configurations 
\cite{Rinaldi:2012vy,Babichev:2013cya,Anabalon:2013oea,Bravo-Gaete:2014haa,Bravo-Gaete:2013dca}, boson and neutron stars \cite{Brihaye:2016lin,Cisterna:2015yla,Cisterna:2016vdx}, Hairy Taub-NUT/Bolt-AdS solutions \cite{Arratia:2020hoy}, as well as holographic applications such that quantum complexity and shear viscosity 
\cite{Feng:2018sqm,Feng:2015oea,Bravo-Gaete:2022lno,Bravo-Gaete:2021hlc,Bravo-Gaete:2020lzs}.}

{On the other hand, through this work the physical system analyzed is based on the model proposed by \cite{Melnikov:2012tb,Fujita:2012fp}. Here, as we will see in the following lines, we start from the same Lagrangian for a Horndeski-Maxwell system, this is (\ref{eq:Lhorn}), together with the Maxwell Lagrangian
\begin{equation}
{\cal L}_{\rm M}=-\dfrac{\kappa}{4e^{2}} F^{\mu \nu} F_{\mu \nu}, \label{L3} 
\end{equation}
where $e$ is a coupling constant and $F_{\mu \nu} = \partial_\mu A_\nu - \partial_\nu A_\mu$ is the Maxwell stress tensor, describing the gravity dual of a field theory on a half-plane.} In the simple plane-symmetric black hole ansatz, we have that only tensionless RS branes are allowed, and that the background solution must be not allowed to model the situation with external electric fields, as in \cite{Fujita:2012fp}. {Even more, as a result of the NBC for the gauge fields, and showing in \cite{Melnikov:2012tb}, the charge density $\rho$ in the dual field theory must be supported by an external magnetic field $B$, where the ratio $\rho/B$, which is equal to the Hall conductivity, is a constant inversely proportional to the coefficients.} In our prescription, this represents the topological terms present in the gravity action: namely, a $m^{2}$ in the bulk action, that is, {an} antisymmetric tensor field $M_{\mu\nu}$ {which} is the effective polarization tensor of the term in the boundary action on the RS branes \cite{Cai:2015bsa,Cai:2014oca,Ghotbabadi:2021mus}. Such behaviors are expected for a quantum Hall system tuned to a quantized value of the conductivity. Furthermore, we provided similar results in the AdS/BCFT holographic model, where, {for example}, we will see how accurately it can account for the physical behaviors expected in a quantum Hall system where, {as was showed before, through AdS/BCFT construction} the Hall conductivity is inversely proportional to the coefficients of the terms that appear in the gravity Lagrangian. {Additionally}, the ratio $\rho/B$ will indicate a localized condensate \cite{Hartnoll:2008kx,Hartnoll:2009sz}.

{Just for completeness, as discussed in \cite{Melnikov:2012tb}, for the classical Hall effect, the charge density and the external magnetic field are independent quantities, that is, the $\rho/B$ ratio depends on} the density of conductance electrons. On the other hand, in the quantum Hall Effect (QHE) the transverse conductivity given by $\sigma_{H}$, {has} plateaus that are independent of either $\rho$ or $B$. These plateaus are generally attributed to disorder \cite{Laughlin:1981jd,Moore:1991ks,Avron85}, {being responsible for the existence of localized electron states} \cite{Melnikov:2012tb}. {Here, the localized states fill the gaps between the Landau levels. Nevertheless, there is no active participation in the Hall conductivity.} 

Finally, we {study} the properties of holographic paramagnetism-ferromagnetism phase transition in the presence of Horndeski gravity {(\ref{eq:Lhorn}). Here, from the matter field part,} we consider the effects of the Maxwell field {(\ref{L3})} on the phase transition of this system, {following \cite{Zhang:2016nvj,Wu:2016uyj}, introducing a massive 2-form coupled field, and neglect the effects of this 2-form field and gauge fields on the background geometry.}  In our analysis, we observe that increasing the strength of parameter $\gamma$, {given in (\ref{eq:Lhorn}), decreases the temperature of the holographic model and leads to a harder formation of the magnetic moment in the black hole background.} {On the other hand, at low temperatures, spontaneous magnetization, and ferromagnetic phase transition happen, but when removes the external magnetic field, this magnetization disappears. As we know, ferromagnetic materials have coercivity, which is the ability to keep their elementary magnets stuck in a certain position. This position can be modified by placing the magnetized material in the presence of an external magnetic field. In this way, a material with high coercivity its elementary magnets resists the change of position. In the material science, experimental framework \cite{Muller2016}, there is a close relationship between the magnetic related to viscosity and coercivity, this relationship was predicted theoretically and observed experimentally. Thus, we have a fundamental role in both cases, that is, between viscosity and coercivity, where they play the so-called activation volume, which is the relevant volume where thermally activated and field-induced magnetization processes occur, respectively. In our work, we will study this way for the paramagnetic material to resist the external magnetic field, through the viscosity/entropy ratio. In our model, this relationship depends on the external magnetic field, the Horndeski parameters, and the boundary size $\Delta\,y_{Q}$ of the RS brane in a non-trivial way.}

{This work is organized as follows: In Section \ref{v1} we consider the gravitational setup, which contains all the information with respect to the AdS$_{4}$/BCFT$_{3}$ duality, showing the solution. Together with the above, in Section \ref{sec:charge} the charge density is obtained for then, in Section \ref{sec:Q} to present the boundary $Q$ profile. In Section \ref{sec:HR}, we perform a holographic renormalization, computing the Euclidean on-shell action, which is related to the free energy
of the corresponding thermodynamic system, where in particular we will focus on the black hole entropy, present in Section \ref{sec:ent}, and the holographic paramagnetism/ferromagnetism phase transition, given in Section \ref{sec:hol}. Finally, Section \ref{v4} is devoted to our conclusions and discussions.}

\section{{Black hole as a probe of AdS/BCFT}}\label{v1}
{As was shown in the introduction, we will present our setup starting} with the total action, which contains all information related to AdS$_{4}$/BCFT$_{3}$ correspondence with probe approximation, so that:
\begin{eqnarray}\label{açao}
S &=&{S^{\mathcal{N}}_{\rm H}+S^{\mathcal{N}}_{\rm M}+S^{\mathcal{N}}_{\rm 2-FF}+S^{\mathcal{N}}_{mat}+S^{Q}_{bdry}+S^{Q}_{mat}+S^{Q}_{ct}},\label{1}
\end{eqnarray}
where  
\begin{equation}\label{eq:actionH-M}
 S^{\mathcal{N}}_{\rm H}=\int_{\mathcal{N}}d^{4} x \sqrt{-g}\; {\cal L}_{\rm H},\qquad  S^{\mathcal{N}}_{\rm M}=\int_{\mathcal{N}}d^{4} x \sqrt{-g}\; {\cal L}_{\rm M},   
\end{equation}
 with  ${\cal L}_{\rm H}$ and ${\cal L}_{\rm M}$ given previously in (\ref{eq:Lhorn})-(\ref{L3}) respectively,
while that 
$S^{\mathcal{N}}_{mat}$ is the action associated to matter sources and:
\begin{eqnarray}
S^{Q}_{bdry}&=&2\kappa\int_{Q}{d^{3}x\sqrt{-h}\mathcal{L}_{bdry}}\nonumber\\
S^{Q}_{mat}&=&2\int_{Q}{d^{3}x\sqrt{-h}\mathcal{L}_{mat}},\nonumber\\
S^{Q}_{ct}&=&2\kappa\int_{ct}{d^{3}x\sqrt{-h}\mathcal{L}_{ct}}\,,
\end{eqnarray} 
with 
{\begin{eqnarray}
&&\mathcal{L}_{bdry}=(K-\Sigma)-\frac{\gamma}{4}(\nabla_{\mu}\phi\nabla_{\nu}\phi n^{\mu}n^{\nu}-(\nabla \phi)^2)K-\frac{\gamma}{4}\nabla_{\mu}\phi\nabla_{\nu}\phi K^{\mu\nu}\,,\label{L5}\\
&&\mathcal{L}_{ct}=c_{0}+c_{1}R+c_{2}R^{ij}R_{ij}+c_{3}R^{2}+b_{1}(\partial_{i}\phi\partial^{i}\phi)^{2}+\cdots,\label{L6}
\end{eqnarray}}
{where in our notations $(\nabla \phi)^2=\nabla_{\mu}\phi\nabla^{\mu}\phi$. In Eq.(\ref{L5}),  $\mathcal{L}_{bdry}$ corresponds to the Gibbons-Hawking $\gamma$-dependent terms associated with the Horndeski gravity (\ref{eq:Lhorn}), where $K_{\mu\nu}=h^{\phantom{\mu}\beta}_{\mu}\nabla_{\beta}n_{\nu}$} is the extrinsic curvature, $K=h^{\mu\nu}K_{\mu\nu}$ is the trace of the extrinsic curvature, $h_{\mu\nu}$ is the induced metric, $n^{\mu}$ is an outward pointing unit normal vector to the boundary of the hypersurface $Q$, $\Sigma$ is the boundary tension on $Q$. { $\mathcal{L}_{mat}$ is the matter Lagrangian on $Q$, while that in}  Eq. (\ref{L6}) ${\cal L}_{ct}$  represents the boundary counterterms, which do not affect the bulk dynamics and will be neglected. 

Following the procedures presented by \cite{Magan:2014dwa,Melnikov:2012tb,Fujita:2011fp,Takayanagi:2011zk,Santos:2021orr} we have imposed the {NBC}:
\begin{eqnarray}
K_{\alpha\beta}-h_{\alpha\beta}(K-\Sigma)-\frac{\gamma}{4}H_{\alpha\beta}=\kappa {\cal S}^{Q}_{\alpha\beta}\,,\label{L7}
\end{eqnarray}
where 
\begin{eqnarray}
&&H_{\alpha\beta}\equiv(\nabla_{\sigma}\phi\nabla_{\rho}\phi\, n^{\sigma}n^{\rho}-(\nabla \phi)^2) (K_{\alpha\beta}-h_{\alpha\beta}K)-(\nabla_{\alpha}\phi\nabla_{\beta}\phi)K\,,\label{L8}\\
&&{\cal S}^{Q}_{\alpha\beta}=-\frac{2}{\sqrt{-h}}\frac{\delta S^{Q}_{mat}}{\delta h^{\alpha\beta}}\,.\label{L9} 
\end{eqnarray}
Considering {the matter stress-energy tensor on $Q$ as a constant (this is ${\cal S}^{Q}_{\alpha\beta}=0$),} we can write
\begin{eqnarray}
K_{\alpha\beta}-h_{\alpha\beta}(K-\Sigma)-\frac{\gamma}{4}H_{\alpha\beta}=0\,.\label{L10}
\end{eqnarray}
{On the other hand, from the gravitational part, given by the Einstein-Horndeski theory and assuming that $S^{\mathcal{N}}_{mat}$ is constant, varying $S^{\mathcal{N}}_{\rm H}$ and $S^{Q}_{bdry}$  with respect to the dynamical fields, we have:
\begin{eqnarray}
{\cal E}_{\alpha\beta}=-\frac{2}{\sqrt{-g}}\frac{\delta S^{\mathcal{N}}}{\delta g^{\alpha\beta}}\,,\quad {\cal E}_{\phi}=-\frac{2}{\sqrt{-g}}\frac{\delta S^{\mathcal{N}}}{\delta\phi} \,,\quad {\cal F}_{\phi}=-\frac{2}{\sqrt{-h}}\frac{\delta S^{Q}_{bdry}}{\delta\phi} \,,\nonumber\\
\end{eqnarray}
where
\begin{eqnarray}
{\cal E}_{\mu\nu}&=&G_{\mu\nu}+\Lambda g_{\mu\nu}-\frac{\alpha}{2}\left(\nabla_{\mu}\phi\nabla_{\nu}\phi-\frac{1}{2}g_{\mu\nu}\nabla_{\lambda}\phi\nabla^{\lambda}\phi\right)\label{11}\nonumber\\
                  &+&\frac{\gamma}{2}\left(\frac{1}{2}\nabla_{\mu}\phi\nabla_{\nu}\phi R-2\nabla_{\lambda}\phi\nabla_{(\mu}\phi R^{\lambda}_{\nu)}-\nabla^{\lambda}\phi\nabla^{\rho}\phi R_{\mu\lambda\nu\rho}\right)\nonumber\\
									&+&\frac{\gamma}{2}\left(-(\nabla_{\mu}\nabla^{\lambda}\phi)(\nabla_{\nu}\nabla_{\lambda}\phi)+(\nabla_{\mu}\nabla_{\nu}\phi)\Box\phi+\frac{1}{2}G_{\mu\nu}(\nabla\phi)^{2}\right)\nonumber\\
									&-&\frac{\gamma g_{\mu\nu}}{2}\left(-\frac{1}{2}(\nabla^{\lambda}\nabla^{\rho}\phi)(\nabla_{\lambda}\nabla_{\rho}\phi)+\frac{1}{2}(\Box\phi)^{2}-(\nabla_{\lambda}\phi\nabla_{\rho}\phi)R^{\lambda\rho}\right),\\
{\cal E}_{\phi}&=&\nabla_{\mu}\left[\left(\alpha g^{\mu\nu}-\gamma G^{\mu\nu}\right)\nabla_{\nu}\phi\right]\,,\label{L11}\\
{\cal F}_{\phi}&=&-\frac{\gamma}{4}(\nabla_{\mu}\nabla_{\nu}\phi n^{\mu}n^{\nu}-(\nabla^{2}\phi))K-\frac{\gamma}{4}(\nabla_{\mu}\nabla_{\nu}\phi)K^{\mu\nu}\,,\label{L12}
\end{eqnarray}
and note that, ${\cal E}_{\phi}={\cal F}_{\phi}$, from the Euler-Lagrange equation.}

{Together with the above, and according to \cite{Rinaldi:2012vy,Babichev:2013cya,Anabalon:2013oea,Bravo-Gaete:2014haa,Bravo-Gaete:2013dca}
, we have a condition that deals to static black hole configurations, avoiding  no-hair theorems \cite{Hui:2012qt}. 
Here, we need to require that the square of the radial component of the conserved current must vanish identically without restricting the radial dependence of the scalar field. Such discussion implies that in Eq. (\ref{L11}):
\begin{equation}
\alpha g_{rr}-\gamma G_{rr}=0\label{L13}\,, 
\end{equation}
and defining $\phi{'}(r)\equiv {\psi}(r)$, where $ (')$ denotes the derivative with respect to $r$, we can show that the equations ${\cal E}_{\phi}=0={\cal E}_{rr}$ are satisfied. In our setup, the four dimensional metric is defined via the following line element
\begin{equation}\label{ansatz}
ds^2= \frac{L^2}{r^2}\left(-f(r)\,dt^2+dx^2+dy^2+\frac{dr^2}{f(r)}\right),
\end{equation}
where $ x_1 \leq x \leq x_2$ and $ y_1 \leq y \leq y_2$, while that from Refs.\cite{Brito:2019ose,Santos:2021orr,Bravo-Gaete:2014haa}, $f(r)$ is the metric function which takes the form 
\begin{eqnarray}
&&f(r)=\frac{\alpha L^{2}}{3\gamma}\left[1-\left(\frac{r}{r_{h}}\right)^{3}\right]\,,\label{L14}
\end{eqnarray}
while that $\psi(r)$ reads
\begin{eqnarray}
&&\psi^{2}(r)=(\phi'(r))^2=-\frac{2L^{2}(\alpha+\gamma\Lambda)}{\alpha\gamma r^{2}f(r)}\,,\label{L15}
\end{eqnarray}
where
\begin{equation}\label{eq:phi}
\phi(r)=\pm{\frac { 2\sqrt {-6(\alpha+\Lambda \gamma)}}{3 \alpha}}\, \tanh^{-1} \left(\sqrt{1-\frac{r^3}{r_h^3}} \right)+\phi_0. \end{equation}
Here, $\phi_0$ and $r_h$ are integration constants, where the last one is related to the location of the event horizon. Following the steps of \cite{Santos:2021orr,Brito:2019ose}, performing the transformations 
\begin{eqnarray}
&&f(r) \rightarrow \frac{\alpha L^{2}}{3\gamma} f(r),\qquad t \rightarrow \frac{3\gamma}{\alpha L^{2}} t,\nonumber\\
&&x \rightarrow \sqrt{\frac{3\gamma}{\alpha L^{2}}} x, \qquad y \rightarrow \sqrt{\frac{3\gamma}{\alpha L^{2}}} y,\qquad L \rightarrow \sqrt{\frac{\alpha}{3\gamma}} L^2,\label{transfor}
\end{eqnarray}
we have that the line element (\ref{ansatz}) is invariant, but now the metric function $f(r)$ takes the form
\begin{eqnarray}
&&f(r)= 1-\left(\frac{r}{r_{h}}\right)^{3}\,\label{L16}
\end{eqnarray}
while the square of the derivative of the scalar field $\psi^2(r)$ takes the form given previously in (\ref{L15}). Here is important to note that from Eqs. (\ref{L15})-(\ref{eq:phi}) we can see that to have a real scalar field,
$$\alpha+\Lambda \gamma \leq 0,$$
where it vanishes when $\alpha=-\Lambda \gamma$.}

{It is important to note that, from the action (\ref{açao}), we can see that there is another contribution, denoted as $S^{\mathcal{N}}_{\rm 2-FF}$, which is responsible to construct the ferromagnetic/paramagnetic model. The above will be explained in the following section.}

\section{The finite charge density}\label{sec:charge}

{As was shown in the previous section, in the action (\ref{açao}) appears the additional contribution 
$$S^{\mathcal{N}}_{\rm 2-FF}=\lambda^2 \int_{\mathcal{N}}d^{4} x \sqrt{-g}\; {\cal L}_{\rm 2-FF},$$
where
\begin{eqnarray}
&&{\cal L}_{\rm 2-FF}=-\frac{1}{12} (dM)^2 - \frac{m^2}{4}M^{\mu \nu} M_{\mu \nu}- \frac{1}{2}M^{\mu \nu}F_{\mu \nu}-\frac{J}{8}V(M).\label{L4}
\end{eqnarray}
Here, the above action defined from the seminal works \cite{Cai:2015bsa,Cai:2014oca}, is coupled through the constant $\lambda$ and constructed via the 2-form $M_{\mu\nu}$, $dM$ is the exterior differential of the 2-form field $M_{\mu\nu}$, this is $(dM)_{\tau \mu \nu} =3 \nabla_{[\tau} M_{\mu \nu]}$ and $(dM)^2=9 \nabla_{[\tau} M_{\mu \nu]} \nabla^{[\tau} M^{\mu \nu]}$, $m$ is a constant related to the mass, while that $V(M)$ describes the self-interaction of polarization tensor, with $J$ a constant, which reads 
\begin{equation}
V(M)=(^{*}M_{\mu \nu} M^{\mu \nu})^2=[^{\ast} (M \wedge M)]^2,\label{L4.1}\\    
\end{equation}
where $(^{*})$ is the Hodge star operator, this is $^{*}M_{\mu \nu}=\frac{1}{2!} \varepsilon^{\alpha \beta}_{\phantom{\alpha \beta} \mu \nu} M_{\alpha \beta}$ and $\varepsilon^{\alpha \beta}_{\phantom{\alpha \beta} \mu \nu}$ is the Levi-Civita Tensor. In the following lines, will restrict our analysis to the probe approximation, that is, from the action Eq. \eqref{açao}, one can derive the corresponding equations of motions for matter fields in the probe approximation, that is, $e^{2}\to\infty$ and $\lambda\to 0$, so that:
\begin{eqnarray}
\nabla^{\mu} \left(F_{\mu \nu}+\frac{\lambda^2}{4}\,M_{\mu\nu} \right) &=& 0,\label{eom1}\\
\nabla^{\tau} (dM)_{\tau \mu \nu} - m^2 M_{\mu \nu} - J(^{\ast}M_{\tau \sigma}M^{\tau \sigma})(^{\ast}M_{\mu \nu}) - F_{\mu \nu} &=&0\,. \label{eom2}
\end{eqnarray}
Given that we are focusing on the probe limit approximation, we are going to disregard any back reaction coming from the two-form field $M_{\mu \nu}$.
In order to analyze the holographic paramagnetism/ferromagnetism and paraelectric/ferroelectric phase transition, we consider the gauge fields $M_{\mu \nu}$ and $A_{\mu}$ we consider the following ansatz:
\begin{eqnarray} 
M_{\mu \nu} &=& -p(r)\,dt\wedge dr + \rho(r)\,dx\wedge dy,\label{Mansatz}  \\
A_{\mu} &=& A_{t}(r)\,dt + B x\,dy, \quad F = dA,\label{Fansatz}
\end{eqnarray}
where $B$ is the external magnetic field.} Using \eqref{ansatz}, \eqref{Mansatz}-\eqref{Fansatz} in the background (\ref{L16}), the field equations \eqref{eom1} and \eqref{eom2} are given by
\begin{eqnarray}
A_{t}'+\left(m^2-\frac{4\,J\,r^4\,\rho^2}{L^{4}}\right)\,p &=& 0,\label{L18} \\
\frac{\rho ''}{L^{2}} +\left(\frac{f'}{f}+\frac{2}{r}\right)\,\frac{\rho'}{L^{2}} - \left(\frac{4\,J\,r^2\,p^2}{fL^{4}}+\frac{m^2}{r^2\,f}\right)\,\rho - \frac{B}{r^2\,f} &=& 0,\label{L19} \\
A_t''+\frac{\lambda^2}{4}\,p' &=& 0\,,\label{L20}
\end{eqnarray}
As we work with probe approximation, the back reaction can be neglected. {Together with the above,} given that the behaviors are asymptotically AdS$_{4}$, we can solve the field equations (\ref{L18})-(\ref{L20}) near {to the boundary (this is $r\to 0$)}. Here, asymptotic solutions are given by
\begin{eqnarray}
&&A_{t}(r)\sim\mu-\sigma r,\\
&&p(r)\sim \frac{\sigma}{m^2},\\
&&\rho(r)\sim\rho_{+}r^{\Delta_{+}}+\rho_{-}r^{\Delta_{-}}-\frac{B}{m^2},\label{eq:rho}\\
&&\Delta_{\pm}=\frac{-1\pm\sqrt{1+4L^2 m^2}}{2}.\label{eq:m}
\end{eqnarray}
{Here,} $\rho_{+}$ and $\rho_{-}$ correspond to the source and vacuum expectation value of the dual operator in the boundary field theory (up to a normalization factor), respectively. {It is worth pointing out that one should take $\rho_{+}=0$, in order to obtain condensation spontaneously \cite{Cai:2015bsa}. From Eq. (\ref{eq:rho}), we can define $\rho_{+}$ and $\rho_{-}$ as
\begin{eqnarray}
&&\rho_{+}\equiv r^{-\Delta_{+}}_{h},\qquad \rho_{-}\equiv r^{-\Delta_{-}}_{h},
\end{eqnarray}
yielding to the asymptotic solution $\rho(r)$ the following structure 
\begin{eqnarray}
\rho(r)\sim \left(\frac{r}{r_{h}}\right)^{\Delta_{+}}+\left(\frac{r}{r_{h}}\right)^{\Delta_{-}}-\frac{B}{m^2}.
\end{eqnarray}
Additionally, and according to \cite{Miao:2018qkc}, we can to analyze the electromagnetic field,  extracted from the four dimensional {\em electromagnetic duality}, in a sense that the theory is invariant under
\begin{eqnarray}
{F}_{\mu\nu}\to ^{*}{F}_{\mu\nu}=\frac{1}{2}\varepsilon_{\mu\nu\alpha\beta} {F}^{\alpha\beta}\label{S3},
\end{eqnarray}
where, as before, $\varepsilon_{\alpha \beta \mu \nu}$ is the Levi-Civita Tensor, transforming the electric field into a magnetic field and vice versa.
Such duality gives that, from the action (\ref{L3}), ${F}_{\mu\nu} {F}^{\mu\nu}=(^{*}{F}_{\mu\nu})(^{*}{F}^{\mu\nu})$, showing that is invariant under (\ref{S3}). Besides, the transformation (\ref{S3}) shows that $\mathcal{F}_{rt}\to (^{*}\mathcal{F}_{rt})=\mathcal{F}_{xy}=\sigma=B$, where $\sigma$ ($B$) is the constant related to the electric (magnetic) field.}
\section{Q-boundary profile}\label{sec:Q}
In this section, we present the boundary $Q$ profile, we assume that $Q$ is parameterized by the equation {$y=y_{Q}(r)$, analyzing the influence of the Horndeski action (\ref{eq:Lhorn}), (\ref{eq:actionH-M}). For this, to find the extrinsic curvature, one has to consider the induced metric on this surface, which reads 
\begin{eqnarray}
ds^{2}_{\rm ind}=\frac{L^{2}}{r^{2}}\left(-f(r)dt^{2}+dx^{2}+\frac{g^{2}(r)dr^{2}}{f(r)}\right)\,,\label{Q1} 
\end{eqnarray}
where $g^{2}(r)=1+{y'}^{2}(r)f(r)$ {and $({'})$ denotes the derivative with respect to the coordinate $r$. Here, the normal vectors on Q are} represented by 
\begin{eqnarray}
n^{\mu}=\frac{r}{Lg(r)}\, \left(0,0,\, 1, \, -{f(r)y{'}(r)}\right)\,.\label{Q2}
\end{eqnarray}
Considering the field equation ${\cal F}_{\phi}=0$ (\ref{L12}), one can solve the Eq. \eqref{L10}, yielding
\begin{eqnarray}
y{'}(r)&=&\frac{(\Sigma L)}{\sqrt{4+\dfrac{\gamma\psi^{2}(r)}{4}-(\Sigma L)^{2}\left(1-\left(\dfrac{r}{r_{h}}\right)^{3}\right)}}\,, 
\end{eqnarray}
\noindent and, with $\psi^2(r)$ given {previously in} Eq. \eqref{L15}, we have
\begin{eqnarray}
y{'}(r)&=&\frac{(\Sigma L)}{\sqrt{4-\dfrac{\xi L^{2}}{2r^{2}\left(1-\left(\dfrac{r}{r_{h}}\right)^{3}\right)}-(\Sigma L)^{2}\left(1-\left(\dfrac{r}{r_{h}}\right)^{3}\right)}}\,,\label{prof}
\end{eqnarray}
where we define
\begin{equation}\label{eq:xi}
\xi=\dfrac{\alpha+\gamma\Lambda}{\alpha}.    
\end{equation}
With all this information, we can plot the $y_{Q}$ profile from Eq. (\ref{prof}), representing the holographic description of BCFT considering the theory (\ref{eq:Lhorn}).
\begin{figure}[!ht]
\begin{center}
\includegraphics[scale=0.12]{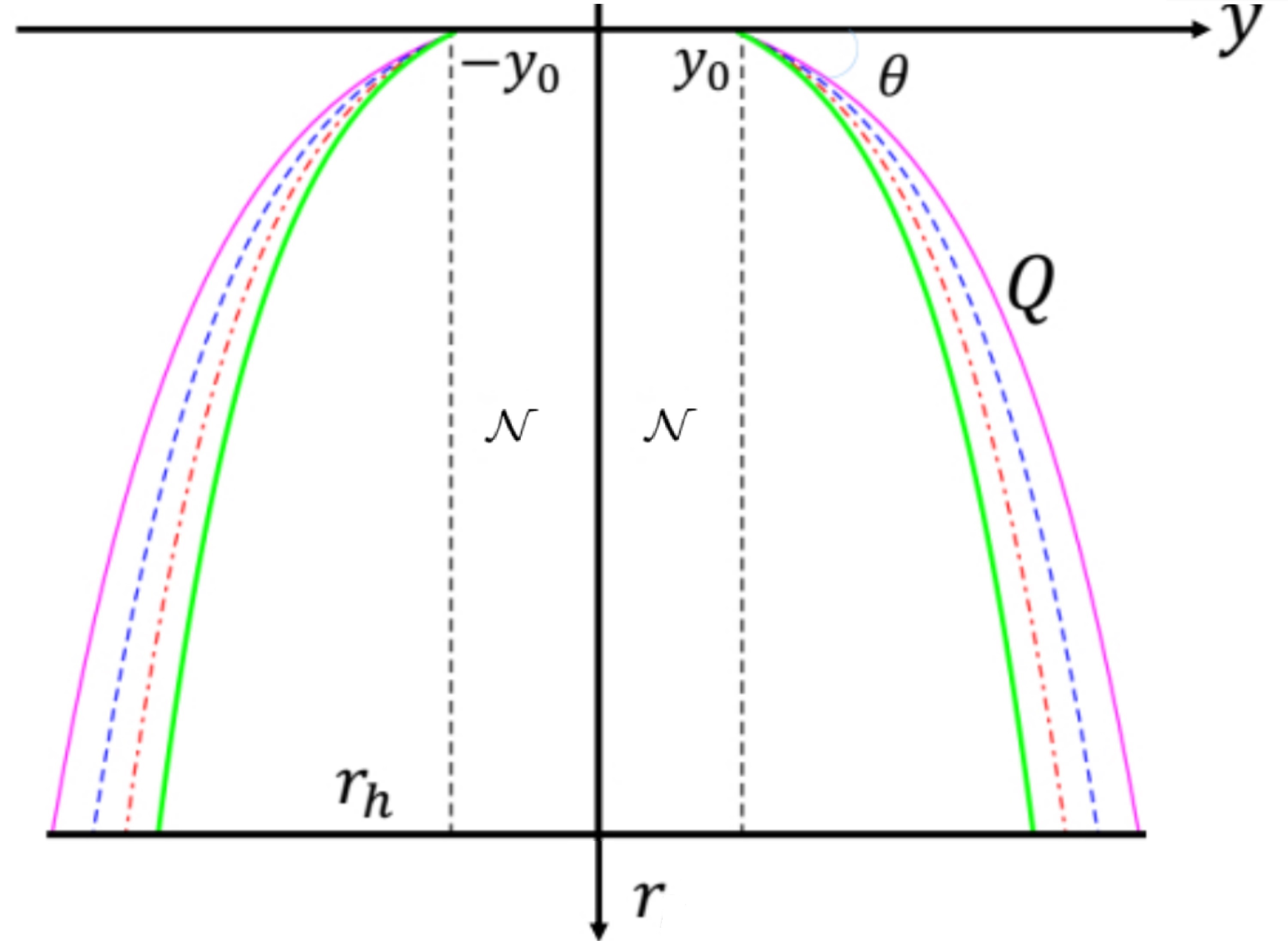}
\caption{{The figure shows the numerical solution for Q boundary profile  from Eq. (\ref{prof})} for the black hole within Horndeski gravity, considering the values for $\theta'=2\pi/3$, $\theta=\pi-\theta'$, $\Lambda=-1$, $\alpha=8/3$ with $\gamma=0$ ({\sl pink curve}), $\gamma=0.1$ ({\sl blue dashed curve }), $\gamma=0.2$ ({\sl red dot dashed} curve), and $\gamma=0.3$ ({\sl green thick curve}). The  dashed parallel vertical lines represent the UV solution, Eq. \eqref{profuv}, that is, Randall-Sundrum branes. The region between the curves Q represents the bulk {$\mathcal{N}$.}}\label{p1}
\end{center}
\end{figure}}

\begin{figure}[!ht]
\begin{center}
\includegraphics[scale=1]{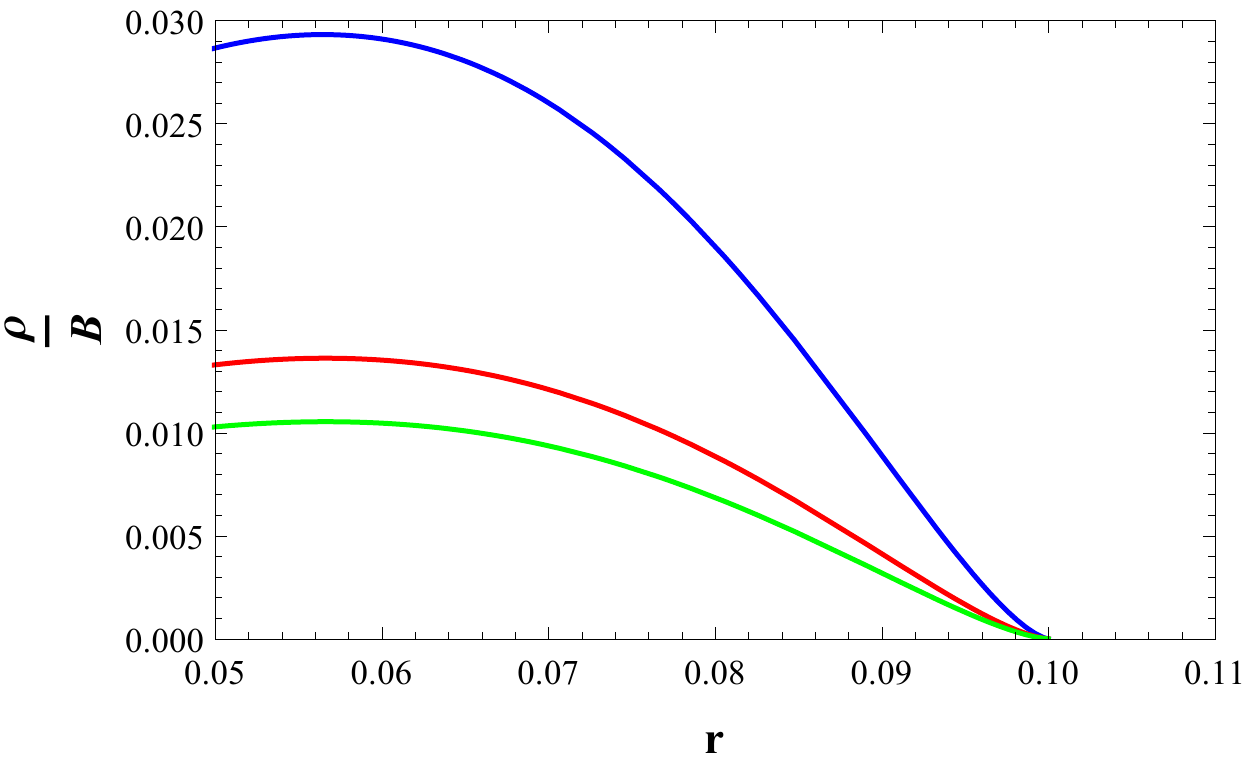}
\caption{{Graphic of the ratio $\rho/B$ with respect to external magnetic field $B$ versus $r$, for different values of the Horndeski parameter $\gamma$. Here, we consider the values $r_{h}=0.1$, $L=1$, $\theta{'}=2\pi/3$, $\Lambda=-1$, $\alpha=0.5$, $m=1$, and $\gamma=1$ (represented through the blue curve), $\gamma=4$ (represented through the red curve), and $\gamma=8$ (represented through the green curve).}}
\label{p0}
\label{ylinhaz}
\end{center}
\end{figure}

{On the other hand, following the steps of \cite{Melnikov:2012tb,dosSantos:2022scy}, we have that the NBC on the gauge field is $n^{\mu}{F}_{\mu\nu}|_{Q}=0$, and $B=\sigma$}. The holographic model (AdS$_{4}$/BCFT$_{3}$) predicts that a constant boundary current in the bulk induces a constant current on the boundary $Q$. Such boundary current can be measured in materials graphene-like. Furthermore, $n^{\mu}M_{\mu\nu}|_{Q}=0$ provide  
\begin{eqnarray}\label{eq:rho/B}
\frac{\rho(r)}{B}=\dfrac{f(r)y'(r)}{m^{2}}.
\end{eqnarray}
{Here, the density $\rho$ and the magnetic field $B$} are no longer two independent parameters. As {the ratio is the Hall conductivity,} this is very reminiscent of the quantum Hall effect (QHE), where this ratio is independent of both $\rho$ and $B$ and is inversely proportional to the topological coefficients, which in our case are the coupling constant $\gamma$ {presents in the Horndeski gravity, together with the parameter from the antisymmetric tensor field $M_{\mu\nu}$, this is $m^2$. In our case, the equation of $y'$ from (\ref{prof}) and then the $\rho/B$ ratio (\ref{eq:rho/B}) can be analyzed by numerical calculations, being represented in  Fig. \ref{p0}. Here, we show  the  ratio $\rho/B$ with respect to external magnetic field $B$ for different values of the Horndeski gravity parameter $\gamma$, where we introduced $\Sigma L=\cos(\theta')$, where $\theta'$ represents the angle between the positive direction of the $y$ axis and $Q$.} At the boundary $Q$, the curves of solutions in the ($\rho,B$) plane will be a localized condensate \cite{Hartnoll:2008kx,Hartnoll:2009sz}.

{Together with the above, in addition to the above numerical solution, we} can analyze some particular cases regarding the study of the UV and IR regimes. Thus, {for the first case,} performing an expansion at $r\to 0$ with, { as before}, $\Sigma L=\cos(\theta{'})$, the equation (\ref{prof}) becomes 
{\begin{eqnarray}
y_{_{UV}}(r)=y_{0}+ \sqrt{\frac{2}{-\xi L^{2}}}\,{r\cos(\theta{'})},
\end{eqnarray}
where $y_0$ is an integration constant.} In the above equation, considering $\xi\to-\infty$, we have
\begin{eqnarray}
y_{_{UV}}(r)=y_{0}={\rm constant}.\label{profuv}
\end{eqnarray}
This is equivalent to keeping $\xi$ finite and a zero tension limit $\Sigma\to 0$, {considering the cases $\theta'=\pi/2$ and $\theta'=3\pi/2$. 
Now, for this regime, we have that the $\rho/B$ ratio takes the form
\begin{eqnarray}
\frac{\rho}{B}=\sqrt{\frac{2}{-\xi L^{2}}}\dfrac{\cos(\theta{'})}{m^{2}}.\label{ratio}
\end{eqnarray}
Here, it turns out a straightforward generalization of a known AdS$_{4}$/CFT$_{3}$ solution, given by the plane-symmetric charged four-dimensional AdS black hole, where only allows for tensionless RS branes in the AdS$_{4}$/BCFT$_{3}$ construction \cite{Melnikov:2012tb}.} In this case, requires that the static uniform charge density is supported by a magnetic field. Specifically, we found that $\rho/B$  is a constant proportional to a ratio of the coefficients appearing in the Horndeski gravity. { These analyses indicate a generalization of the AdS$_{4}$ black hole can describe a quantum Hall system at a plateau of the transverse conductivity. Additionally,} the AdS/BCFT setup yields that the Hall conductivity is inversely proportional to a sum of the coefficients of the topological terms appearing in the gravity Lagrangian. That is, we obtain that $\sigma_{H}=\rho/B$, which from the equation (\ref{ratio})
{\begin{eqnarray}
\sigma_{H}=\sqrt{\frac{2}{-\xi L^{2}}} \dfrac{\cos(\theta{'})}{m^{2}},\label{Hall1}
\end{eqnarray}
where, as was shown in the introduction, in QHE the conductivity is related to the number of filled Landau levels (filling fraction), namely, by
\begin{eqnarray}
\dfrac{h}{e^{2}}\sigma_{H}=\sqrt{\frac{2}{-\xi L^{2}}} \dfrac{\cos(\theta{'})}{m^{2}},
\end{eqnarray}}
where $e^{2}/h$ is the magnetic flux quantum. In this way, the holographic description seems to provide results similar to the description of the QHE {obtained in } \cite{Moore:1991ks,Avron85}.  In our case, we have an extension of the covariant form of the Hall relation $\rho=\sigma_{H}B$.

For the IR case, we take  $r\to +\infty$ so that from Eq. \eqref{prof} implies  that $\lim_{r\to +\infty} (\phi'(r))^2=0$, and then $\phi=$ constant, which ensures a genuine vacuum solution. Plugging this result in Eq. (\ref{prof}), in the limit $r\to\infty$, we have
{\begin{eqnarray}
y'_{_{IR}}(r)\sim \left(\frac{r_h}{r}\right)^{3/2}+O\left(\frac{1}{r^{2}}\right),
\end{eqnarray}
and $y^{'}_{_{IR}}(r) \to 0$ when $r \to +\infty$, which implies from (\ref{ratio}) that $\rho/B\to 0$}. Such value {becomes the on-shell} action finite.

{For the sake of completeness,} an approximate analytical solution for $y(r)$ can be obtained by performing an expansion for $\xi$ very small from Eq. (\ref{prof}), {this is
$$y'_{Q}=\frac{\cos(\theta')}{\sqrt{
4-\cos^2(\theta')f(r)}}+\frac{L^2
\cos(\theta) \xi}{4 r^2 f(r) (4-\cos^2(\theta')f(r))^{3/2}}+ O(\xi^2),$$
with $f$ given previously in (\ref{L16}), and considering} this expansion up to the first order, we obtain
{\begin{eqnarray}
&&y_{Q}(r)=y_{0}+\frac{r\cos(\theta{'})}{\sqrt{ (r^3-r_h^3)\cos^2(\theta{'})+4 r_h^3}}\,\sqrt{\frac{ 4 r_h^3-(r^3-r_h^3)\cos(2\theta{'})}{4-\cos^2(\theta{'})}}\nonumber\\
&&\times _2F_1\left(\frac{1}{3},\frac{1}{2};\frac{4}{3};-\frac{r^3\cos^2(\theta{'})}{r^{3}_{h} (4-\cos^2(\theta{'}))}\right)+\xi \int \frac{L^2
\cos(\theta)}{4 r^2 f(r) (4-\cos^2(\theta')f(r))^{3/2}} \, dr+O(\xi^2),\label{prof1}
\end{eqnarray}}
{where $_2F_1(a,b;c;x)$ is the hypergeometric function.} 

\section{Holographic renormalization}\label{sec:HR}
In our setup, we will compute the Euclidean on-shell action, which is related to the free energy of the corresponding  thermodynamic system. Thus, our holographic scheme takes into account the contributions of AdS$_{4}$/BCFT$_{3}$ correspondence within Horndeski gravity. Let us start with the Euclidean action given by $I_{E}=I_{bulk}+2I_{bdry}$, i.e.,

\begin{eqnarray}
&&{I_{bulk}=-\frac{1}{16\pi G_N}\int_{\mathcal{N}}{\sqrt{g}d^{4}x\left(R-2\Lambda+\frac{\gamma}{2} G_{\mu \nu} \nabla^{\mu} \phi \nabla^{\nu} \phi\right)}-\frac{1}{8\pi G_N}\int_{\mathcal{M}}{d^{3}x\sqrt{\bar{\gamma}}\mathcal{L}_{\mathcal{M}}},}\\
&&{\mathcal{L}_{\mathcal{M}}=K^{({\bar{\gamma}})}-\Sigma^{(\bar{\gamma})}-\frac{\gamma}{4}(\nabla_{\mu}\phi\nabla_{\nu}\phi n^{\mu}n^{\nu}-(\nabla\phi)^{2})K^{(\bar{\gamma})}-\frac{\gamma}{4}\nabla^{\mu}\phi\nabla^{\nu}\phi K^{(\bar{\gamma})}_{\mu\nu}.}
\end{eqnarray}
Together with the above, $g$ is the determinant of the metric $g_{\mu\nu}$ on the bulk $\mathcal{N}$, {while that} $\bar{\gamma}$ is the induced metric,  the surface tension on $\mathcal{M}$ {is represented} with $\Sigma^{(\bar{\gamma})}$, {and $K^{({\bar{\gamma}})}$ corresponds to the extrinsic curvature on $\mathcal{M}$}. On the other hand, for the boundary, {we have the following expressions}
\begin{eqnarray}
&&I_{bdry}=-\frac{1}{16\pi G_{N}}\int_{\mathcal{N}}{\sqrt{g}d^{4}x {\left(R-2\Lambda+\frac{\gamma}{2} G_{\mu \nu} \nabla^{\mu} \phi \nabla^{\nu} \phi\right)}}-\frac{1}{8\pi G_N}\int_{Q}{d^{3}x\sqrt{h}\mathcal{L}_{bdry}},\\
&&\mathcal{L}_{bdry}=(K-\Sigma)-\frac{\gamma}{4}(\nabla_{\mu}\phi\nabla_{\nu}\phi n^{\mu}n^{\nu}-(\nabla\phi)^{2})K-\frac{\gamma}{4}\nabla^{\mu}\phi\nabla^{\nu}\phi K_{\mu\nu}.
\end{eqnarray}
Thus, in order to compute the bulk action $I_{bulk}$, we consider the induced metric on { the bulk, which is obtained from (\ref{ansatz}) after the transformation $\tau=i t$, given by
\begin{eqnarray}
ds^{2}_{ind}=\bar{\gamma}_{\mu\nu}dx^{\mu}dx^{\nu}=\frac{L^{2}}{r^{2}}\left(f(r)d\tau^{2}+dx^{2}+dy^{2}+\frac{dr^{2}}{f(r)}\right).\label{mett1}
\end{eqnarray}
Here, we have that $0 \leq \tau \leq \beta$, where from Eq. (\ref{L16})
\begin{eqnarray}\label{eq:Th}
\beta=\frac{1}{T}=\left(\frac{|f'(r_h)|}{4 \pi}\right)^{-1}=\frac{4 \pi r_h}{3},
\end{eqnarray}
where $T$ is the Hawking Temperature, the above allows us to obtain the following quantities: 
$$R=-\frac{12}{L^{2}},\qquad \Lambda=-\frac{3}{L^{2}},\qquad K^{({\bar{\gamma}})}=\frac{3}{L},\qquad \Sigma^{(\bar{\gamma})}=\frac{2}{L}.$$} 
Thus, we have all elements needed to construct the bulk action $I_{bulk}$. {In the process of holographic renormalization, we need to introduce a cutoff $\epsilon$ to remove the IR divergence on the bulk side and we can provide that:}
{
\begin{eqnarray}
&&I_{bulk}=\frac{1}{16\pi G_{N}}\int{d^{2}x}\int^{\frac{4 \pi r_h}{3}}_{0}{d\tau}\int^{r_{h}}_{\epsilon}{dr\sqrt{g}\left(R-2\Lambda+\frac{\gamma}{2}G^{rr}\psi^{2}(r)\right)}\nonumber\\
&&+\frac{1}{16\pi G_{N}}\int{d^{2}x}\int^{\frac{4 \pi r_h}{3}}_{0}{d\tau}{\frac{L^{2}\sqrt{f(\epsilon)}}{\epsilon^{3}}},\\
&&I_{bulk}=-\frac{L^{2}V}{8r^{2}_{h}G}\left(1-\frac{\xi}{4}\right),\label{eq:bulk}
\end{eqnarray}
with $\xi$ given previously in (\ref{eq:bdry1}) and, in our notations, $V=\int{d^{2}x}=\Delta{x} \Delta{y}=(x_2-x_1)(y_2-y_1)$.}
Now, computing the $I_{bdry}$, {we introduce a cutoff $\epsilon$ to remove the UV divergence on the boundary side, and with this information, we have:}
\begin{eqnarray}
I_{bdry}=\frac{r_{h}L^{2}{\Delta y_{Q}}}{2G_{N}}\left(1-\frac{\xi}{4}\right)\int^{r_{h}}_{\epsilon}{\frac{\Delta y_{_Q}(r)}{r^{4}}dr}-\frac{r_{h}L^{2}\sec(\theta{'}){\Delta y_{Q}}}{2G_{N}}\int^{r_{h}}_{\epsilon}{\frac{\Delta y_{_Q}(r)}{r^{3}}dr}\label{Idry}
\end{eqnarray}
Here, $\Delta y_{Q}$ is a constant and {$\Delta y_{Q}(r):=y_{Q}(r)-y_0$ is obtained from the equation (\ref{prof1}).} As we know, from the point of view of AdS/CFT correspondence, IR divergences in AdS correspond to UV divergences in CFT. This relationship is known as the IR-UV connection. {Thus, based on this duality, we can reduce the above equation (\ref{Idry}) after some eliminations of terms that produce divergences to the following form:}
\begin{eqnarray}
&&I_{bdry}=-\frac{L^{2}\Delta\,y_{Q}}{2 G_{N}}\left(1-\frac{\xi}{4}\right)\left(\frac{\xi\,L^{2}b(\theta{'})}{5r^{4}_{h}}+\frac{q(\theta^{'})}{4r^{2}_{h}}\right)\nonumber\\
&&+\frac{L^{2}\sec(\theta{'})\Delta\,y_{Q}}{2G_{N}}\left(\frac{\xi\,L^{2}b(\theta{'})}{4r^{3}_{h}}+\frac{q(\theta^{'})}{2r_{h}}\right),\label{eq:bdry1}
\end{eqnarray}
\noindent where
\begin{eqnarray}
b(\theta{'})=\frac{\cos(\theta{'})}{4(4-\cos^{2}(\theta{'}))^{3/2}},\qquad  
q(\theta{'})=\frac{\cos(\theta{'})}{\sqrt{4-\cos^{2}(\theta{'})}}\,\label{eq:bdry2}. 
\end{eqnarray}
With all the above information, from Eqs. (\ref{eq:bulk}) and (\ref{eq:bdry1})-(\ref{eq:bdry2}), we can compute $I_{E}=I_{bulk}+2I_{bdry}$ as:

\begin{eqnarray}
&&I_{E}=-\frac{L^{2}V}{8r^{2}_{h}G_N}\left(1-\frac{\xi}{4}\right)-\frac{L^{2}\Delta\,y_{Q}}{G_{N}}\left(1-\frac{\xi}{4}\right)\left(\frac{\xi\,L^{2}b(\theta{'})}{5r^{4}_{h}}+\frac{q(\theta^{'})}{4r^{2}_{h}}\right)\nonumber\\
&&+\frac{L^{2}\sec(\theta{'})\Delta\,y_{Q}}{G_{N}}\left(\frac{\xi\,L^{2}b(\theta{'})}{4r^{3}_{h}}+\frac{q(\theta^{'})}{2r_{h}}\right)\label{freeEBH}.
\end{eqnarray}
{Here, $I_{E}$ is the approximated analytical expression for the Euclidean action. This equation is essential to construct the free energy and extract all thermodynamic quantities in our setup, as we show in the next section.}

%

\section{Black hole entropy}\label{sec:ent}

Now, we will compute the entropy related to the black hole considering the contributions of the AdS/BCFT correspondence in the Horndeski gravity. Free energy is defined as
\begin{equation}\label{FE}
 \Omega=T I_E \,,  
\end{equation}
one can obtain the corresponding entropy as:
\begin{eqnarray}
S=-\frac{\partial\,\Omega}{\partial T}\,\label{BT7}
\end{eqnarray}
{where $T$ is the Hawking Temperature}. By plugging the Euclidean {\em on-shell action} $I_E$ from Eq.\eqref{freeEBH}, {and replacing $T$ obtained previously in (\ref{eq:Th}), we have
\begin{eqnarray}\label{eq:ent-total}
S_{\rm total}&=&S_{\rm bulk}+S_{\rm bdry},
\end{eqnarray}
where
\begin{eqnarray}
S_{\rm bulk}&=&\frac{L^{2}V}{4r^{2}_{h}G_{N}}\left(1-\frac{\xi}{4}\right),\label{eq:Entbulk}\\
S_{\rm bdry}&=&\frac{L^{2}\Delta\,y_{Q}}{G_{N}}\left(1-\frac{\xi}{4}\right)\left(\frac{\xi\,L^{2}b(\theta{'})}{5r^{4}_{h}}+\frac{q(\theta^{'})}{4r^{2}_{h}}\right)\nonumber\\
&-&\frac{L^{2}\sec(\theta{'})\Delta\,y_{Q}}{G_{N}}\left(\frac{\xi\,L^{2}b(\theta{'})}{4r^{3}_{h}}+\frac{q(\theta^{'})}{2r_{h}}\right).\label{BT8}
\end{eqnarray}
The interpretation for this total entropy can be identified with the Bekenstein-Hawking formula for the black hole: 
\begin{eqnarray}
S_{BH}=\frac{A}{4G_{N}}\label{BT9}\,,
\end{eqnarray}
where, in this case  
\begin{eqnarray}
A&=&\frac{L^{2}V}{2r^{2}_{h}}\left(1-\frac{\xi}{4}\right)+4L^{2}\Delta\,y_{Q}\left(1-\frac{\xi}{4}\right)\left(\frac{\xi\,L^{2}b(\theta{'})}{5r^{4}_{h}}+\frac{q(\theta^{'})}{4r^{2}_{h}}\right)\nonumber\\
&&-4L^{2}\sec(\theta{'})\Delta\,y_{Q}\left(\frac{\xi\,L^{2}b(\theta{'})}{4r^{3}_{h}}+\frac{q(\theta^{'})}{2r_{h}}\right).\,\label{BT10}
\end{eqnarray}
Here, $A$ is the total area of the AdS black hole in the Horndeski contribution terms for the bulk and the boundary $Q$. We can see that the information is bounded by the black hole area. Then, the equation \eqref{BT10} suggests that the information storage increases with increasing $|\xi|$, as long as $\xi<0$. 

Together with the above, with respect to the boundary contribution of  (\ref{BT8}), we have that this expression is the entropy of the BCFT corrected by the Horndeski terms parametrized by $\xi$, given previously in (\ref{eq:xi}). In this case, the results presented in Refs. \cite{Melnikov:2012tb,Magan:2014dwa} are recovered in the limit $\xi\to 0$. Besides, still analyzing Eq. \eqref{BT8}, due to the effects of the Horndeski gravity, there is a non-zero boundary entropy even if we consider the zero temperature scenario, similar to an extreme black hole. This can be seen if one takes the limit $T\to 0$ (or $r_h \to \infty$) in Eq.\eqref{BT8}, then we do not get the denominated residual boundary entropy, as discussed in \cite{Santos:2021orr}.

{On the other hand, through Eq. (\ref{ratio}) we have
\begin{eqnarray}
S^{magnetic}_{bdry}&=&\frac{L^{2}\Delta\,y_{Q}}{G_{N}}\left(1-\frac{\xi}{4}\right)\left(-\frac{2B^{2}\cos^{2}(\theta^{'})}{m^{2}\rho^{2}}\frac{b(\theta{'})}{5r^{4}_{h}}+\frac{q(\theta^{'})}{4r^{2}_{h}}\right)\nonumber\\
&-&\frac{L^{2}\sec(\theta{'})\Delta\,y_{Q}}{G_{N}}\left(-\frac{2B^{2}\cos^{2}(\theta^{'})}{m^{2}\rho^{2}}\frac{b(\theta{'})}{4r^{3}_{h}}+\frac{q(\theta^{'})}{2r_{h}}\right),\label{BT11ext}
\end{eqnarray}
where $m^{2}>-1/(4L^2)$. For the entropy bound, the restriction on $m^{2}$ comes from Eq. (\ref{eq:m}). A well-defined probe limit demands that the charge density contributed by the polarization should be finite. At low temperatures, below the critical, in the ferromagnetic region, we can observe that our entropy is $S_{magnetic}^{bdry}\propto B^{2}$, that is, has a square dependence on the external magnetic field and this is a characteristic of ferromagnetic systems. Furthermore, we can observe that $S^{magnetic}_{bdry}$ is the magnetic entropy of the boundary $Q$, and we can observe that for ferromagnetic materials, the magnetic entropy is associated with the disorder of the magnetic moments. In addition, these materials have spontaneous magnetization. So when we remove the applied magnetic field, they still show magnetization.
\section{Holographic paramagnetism/ferromagnetism phase transition}\label{sec:hol}
In this section, we present the holographic paramagnetism/ferromagnetism phase transition through the boundary contribution from the entropy (\ref{BT11ext}). For this, we start considering the free energy $\Omega$ from (\ref{freeEBH}) -(\ref{FE}), where the first law of black holes thermodynamics, considering the canonical ensemble, takes the form
\begin{eqnarray}\label{eq:first-law}
d\Omega=-P dV-S dT,
\end{eqnarray}
where, in addition to the entropy $S$ as well as the Hawking temperature $T$, the pressure $P$ and the volume $V$ appear, yielding
$$\Omega=\epsilon-TS,$$
where $\epsilon$ takes the role of the energy density.

As a first thermodynamic quantity to study, we will consider the entropy $S$, from Eq. (\ref{eq:ent-total}), calculated in the previous section, and represented graphically in Fig. \ref{ST}, with respect to the Hawking temperature $T$ (\ref{eq:Th}). Here, in the right panel (left panel) there is (not) an external magnetic field $B$. Concretely,  we see that the right panel exhibit similar behavior as analyzed in \cite{PhysRevB.72.024403}, as for example ferromagnetic materials with nearly zero coercivity and hysteresis. On the other hand, in the left panel, when the external magnetic field is removed (this is $B=0$), we still have a disorder of magnetic moments, this is a characteristic of paramagnetism.

\begin{figure}[!ht]
\begin{center}
\includegraphics[scale=0.71]{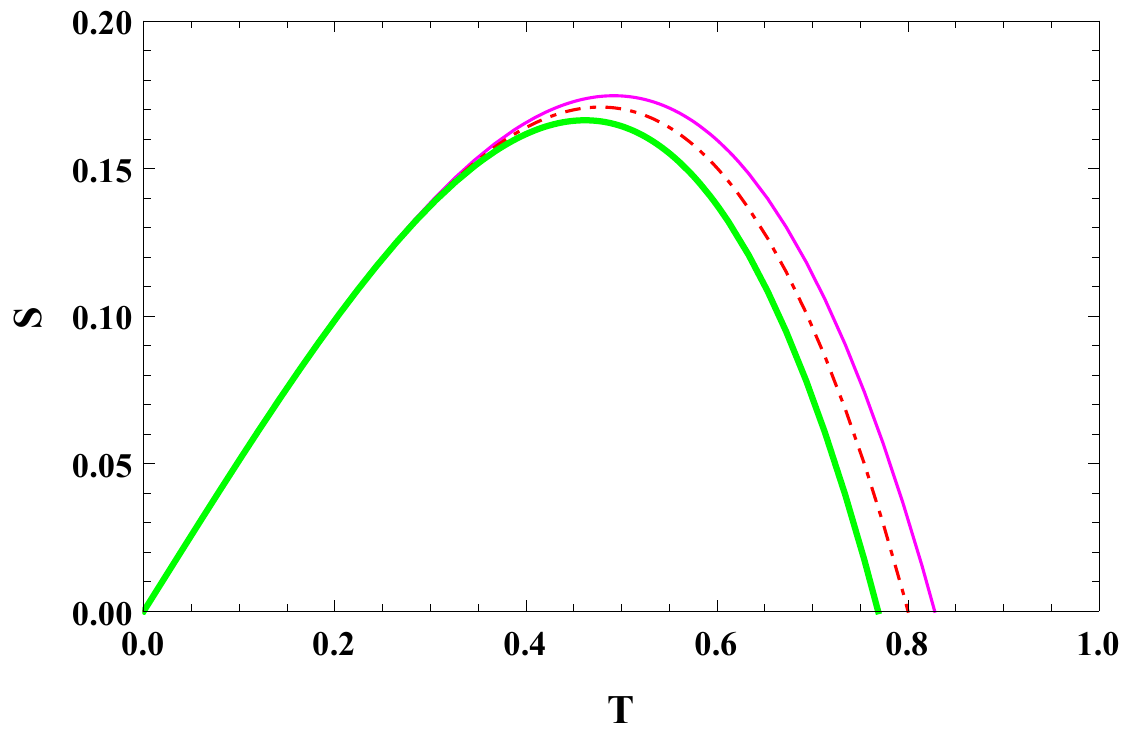}
\includegraphics[scale=0.71]{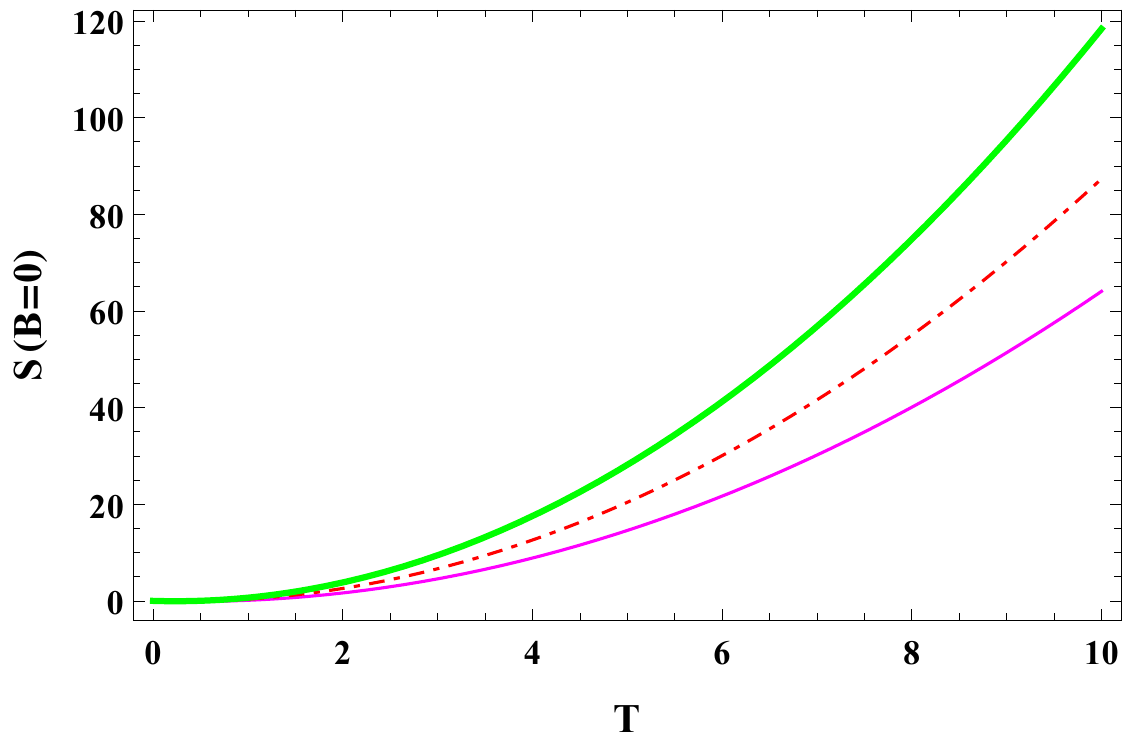}
\caption{{\sl Right panel:} The behavior of the entropy $S$ with the temperature $T$ with different values for $\alpha=8/3$, $m=1/8$, $B=(4/5)T$, $\rho=1/4$, $\Lambda=-1$, $V=1$, $G_N=1$, $\theta{'}=2\pi/3$ with $\gamma=1$ ({\sl pink curve}), $\gamma=4$ ({\sl red dot dashed  curve}), $\gamma=8$ ({\sl  green thick curve}). {\sl Left panel:} The behavior of the entropy $S$ with respect the temperature $T$, with different values for $B=0$.}\label{p3} 
\label{ST}
\end{center}
\end{figure}

The second parameter that we analyze is the heat capacity $C_{V}$, which allows us to analyze local thermodynamic stability, defined in the following form  
\begin{equation}
     \begin{gathered}
     C_{V}=T\bigg(\frac{\partial S}{\partial T}\bigg)_{V}=-T\bigg(\frac{\partial^{2} \Omega}{\partial T^{2}}\bigg)_{V},\label{eq:QUANT.1}
    \end{gathered}
\end{equation}
where the sub-index $V$ from Eq. (\ref{eq:QUANT.1}) represents at volume constant. From Fig. \ref{p4}, we can see that in the right panel, the black hole can switch between stable ($C_V>0$) and unstable ($C_V<0$) phases, depending on the sign of heat capacity $C_V$. This phase transition occurs, due to the spontaneous electric polarization, which was realized in our model from the application of the magnetic external field. Moreover, in the region $C_V>0$, we have structures built like magnetic domes on the boundary $Q$. Additionally, in Fig. \ref{p4}, one can see the influence of Horndeski gravity (represented via the constant $\gamma$) with respect to the temperature $T$, where the phase transition occurs for some ranges of values for $T$ when the external magnetic field is null, that is, $B=0$.

\begin{figure}[!ht]
\begin{center}
\includegraphics[scale=0.7]{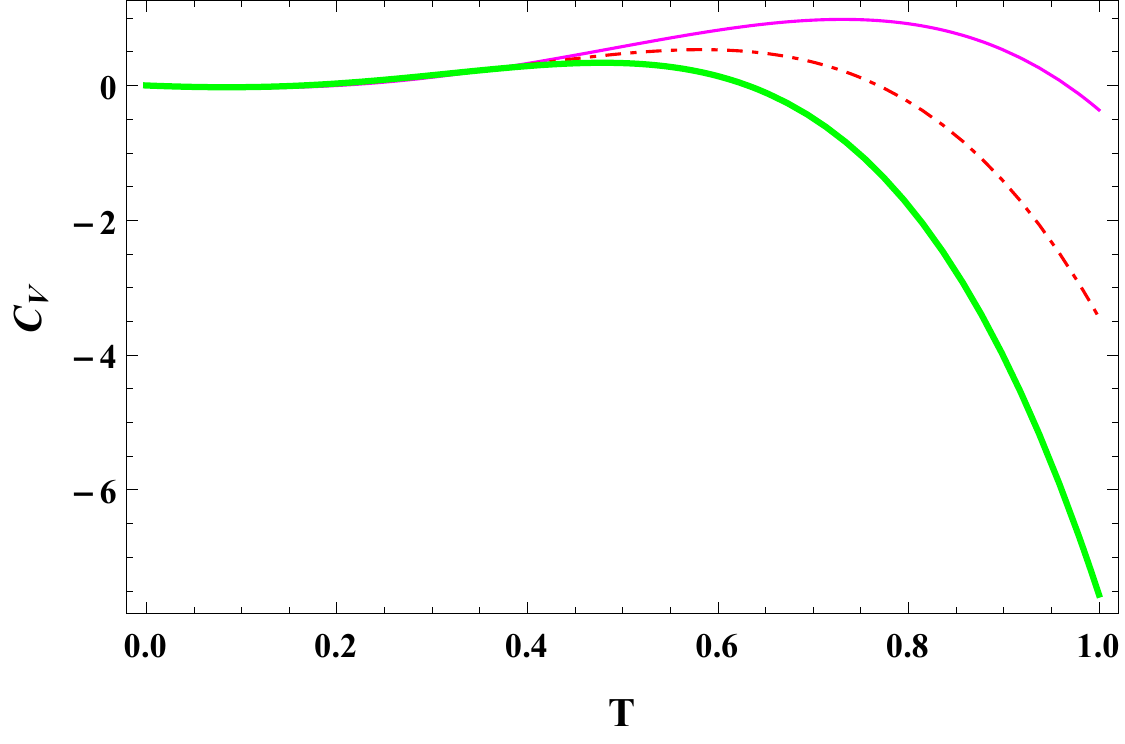}
\includegraphics[scale=0.7]{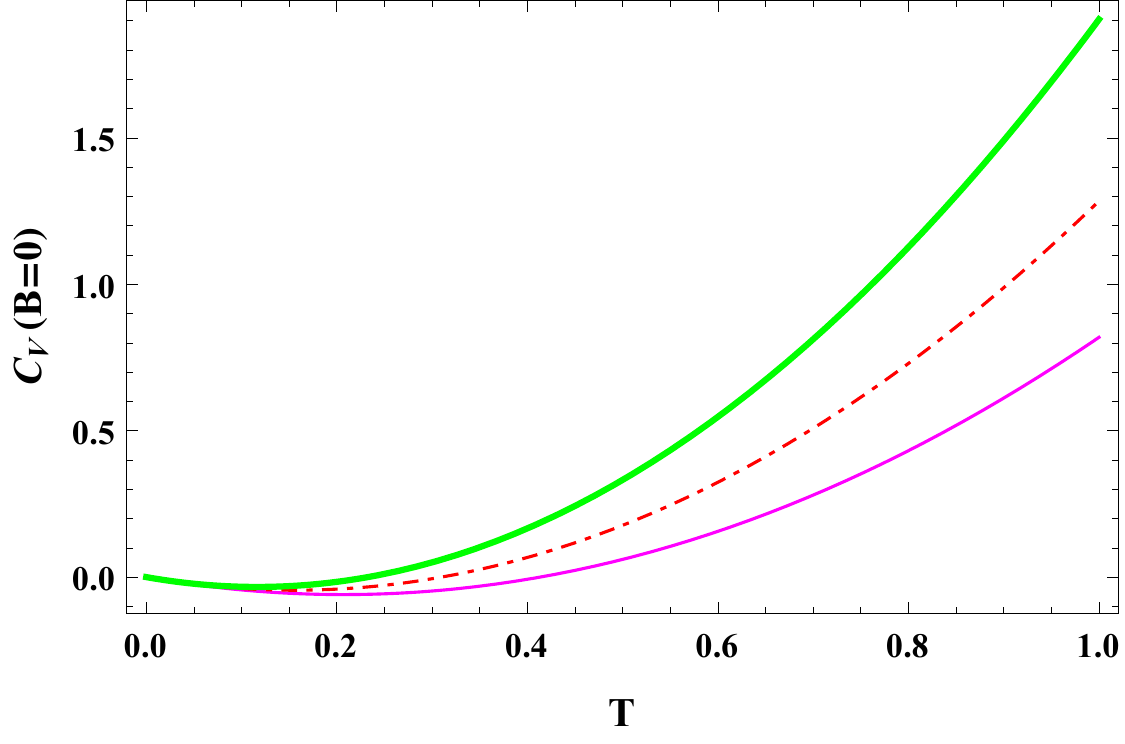}
\caption{{\sl Right panel:} The behavior of the heat capacity $C_{V}$ with the temperature $T$ with different values for $\alpha=8/3$, $m=1/8$, $B=(4/5)T$, $\rho=1/4$, $\Lambda=-1$, $\theta{'}=2\pi/3$ with $\gamma=1$ ({\sl pink curve}), $\gamma=4$ ({\sl red dot dashed  curve}), $\gamma=8$ ({\sl green thick  curve}). {\sl Left panel:} The behavior of the heat capacity $C_{V}$ with respect the temperature $T$, with different values for $B=0$.}\label{p4}
\end{center}
\end{figure}
Additionally, we can obtain the heat capacity at constant pressure $C_P$, which reads
\begin{eqnarray}\label{eq:CP}
C_{P}=T\bigg(\frac{\partial S}{\partial T}\bigg)_{P},
\end{eqnarray}
and, from Fig. \ref{pp1}, we can see that in the right panel, the black hole can switch between stable ($C_P>0$), describing a ferromagnetic material, and unstable ($C_P<0$), describing a paramagnetic material, depending on the sign of heat capacity. This phase transition occurs, as in the previous case, due to spontaneous electric polarization. Moreover, in the region $C_P>0$, we have structures built like magnetic domes on the boundary $Q$, wherein the experimental specific frame, these heat curves without magnetic field can represent a material like $DyAl2$ \cite{PhysRevB.72.024403}. On the other hand, the left panel represents the heat capacity $C_P$ where $B=0$, where we can see, that is locally unstable ($C_P<0$).
\begin{figure}[!ht]
\begin{center}
\includegraphics[scale=0.71]{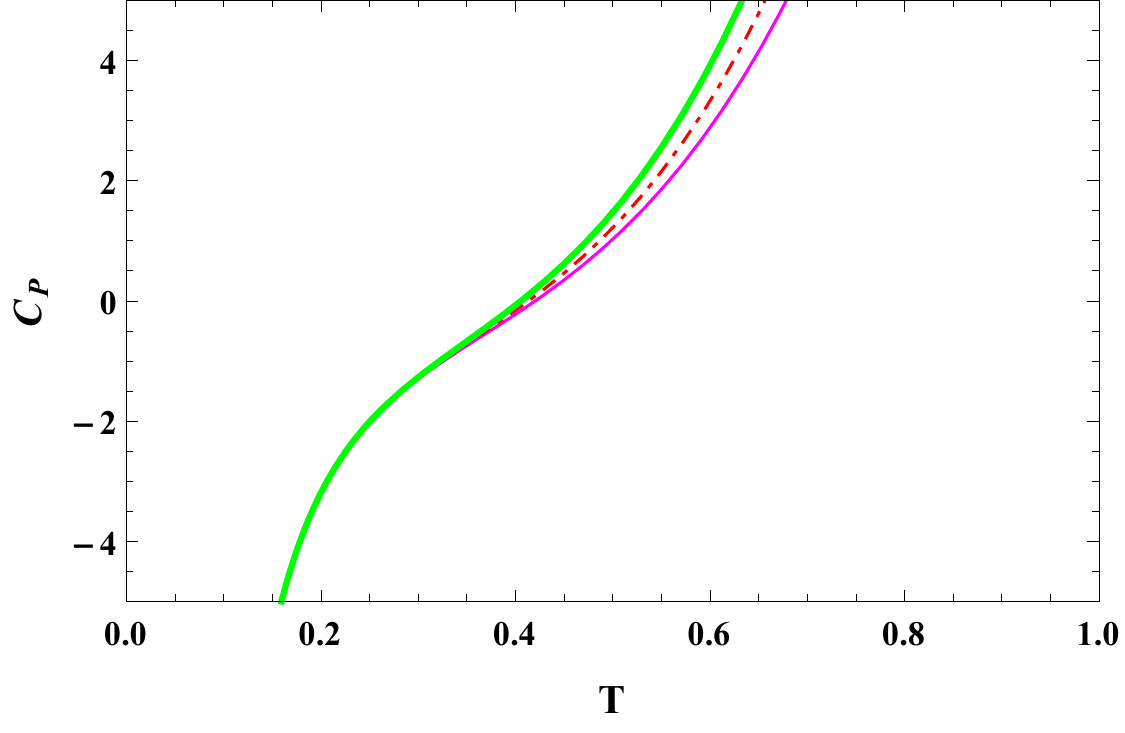}
\includegraphics[scale=0.71]{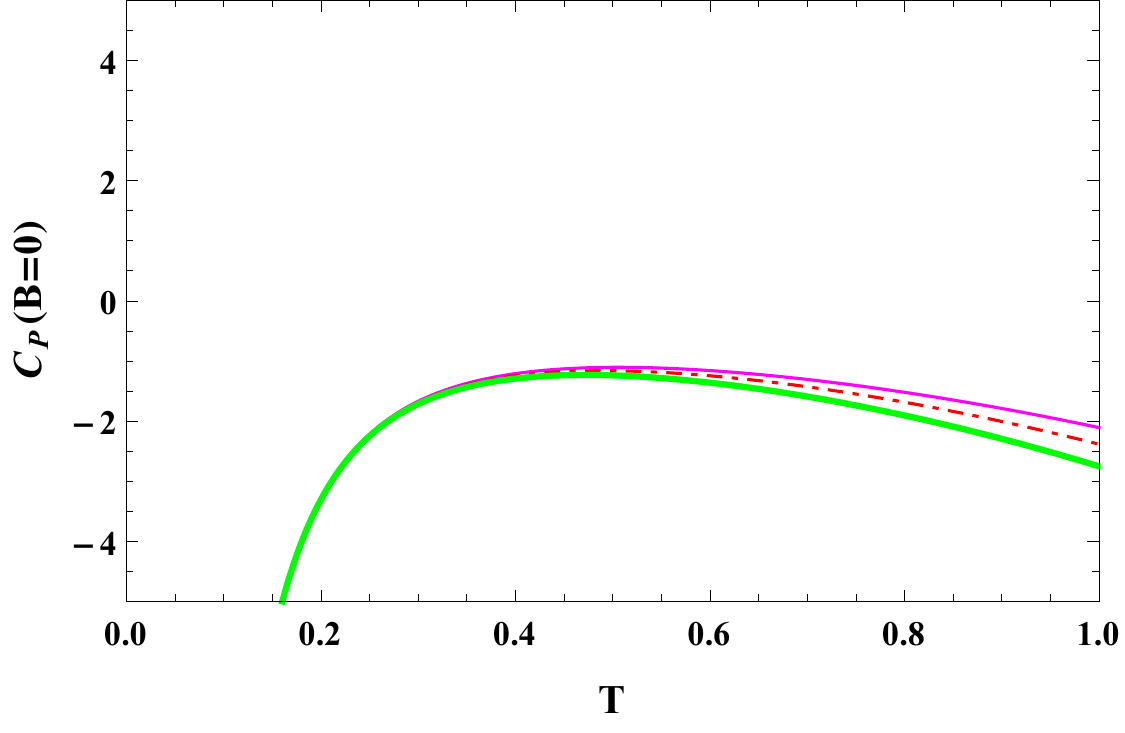}
\caption{{\sl Right panel:} The behavior of the $C_{P}$ with respect to the temperature $T$ with different values for $\alpha=8/3$, $m=1/8$, $B=(4/5)T$, $\rho=1/4$, $\Lambda=-1$, $\theta{'}=2\pi/3$ with $\gamma=1$ ({\sl pink curve}), $\gamma=4$ ({\sl red dot dashed curve}), $\gamma=8$ ({\sl green thick curve}). {\sl Left panel:} The behavior of $C_{P}$ with respect $T$, with different values for $B=0$.}\label{pp1}
\end{center}
\end{figure}

{Additionally, we can derive other quantities, as for example the magnetization density $m$, and magnetic susceptibility $\chi$, following the steps of \cite{Hartnoll:2009sz}, given by} 
{\begin{eqnarray}
&&m=-\bigg(\frac{\partial\,\Omega}{\partial B}\bigg),\quad\chi=\bigg(\frac{\partial^{2} \Omega}{\partial B^{2}}\bigg)\nonumber\\
&&m=\frac{L^{2}\Delta\,y_{Q}T}{G_{N}}\left(1-\frac{\xi}{4}\right)\left(\frac{4B\cos^{2}(\theta^{'})}{m^{2}\rho^{2}}\frac{b(\theta{'})}{5r^{4}_{h}}\right)-\frac{L^{2}\sec(\theta{'})\Delta\,y_{Q}T}{G_{N}}\left(\frac{B\cos(\theta^{'})}{m^{2}\rho^{2}}\frac{b(\theta{'})}{4r^{3}_{h}}\right),\label{mag2}\\
&&\chi=-\frac{L^{2}\Delta\,y_{Q}T}{G_{N}}\left(1-\frac{\xi}{4}\right)\left(\frac{4\cos^{2}(\theta^{'})}{m^{2}\rho^{2}}\frac{b(\theta{'})}{5r^{4}_{h}}\right)+\frac{L^{2}\sec(\theta{'})\Delta\,y_{Q}T}{G_{N}}\left(\frac{\cos(\theta^{'})}{m^{2}\rho^{2}}\frac{b(\theta{'})}{4r^{3}_{h}}\right).\label{mag1}
\end{eqnarray}
As we can see from equations (\ref{mag2}) and (\ref{mag1}) that $m=-\chi\,B$, the RS brane behaves like a paramagnetism material, that is, when we remove the external magnetic field, the equation (\ref{mag2}) disappears and the entropy linked disorder increases, as shown in Fig. \ref{p3}. On the other hand, from the equation (\ref{mag1}), the susceptibility is not null for zero magnetic fields (this is $B=0$). Thus, we can conclude that paramagnetic materials have a low coercivity, that is, their ability to remain magnetized is very low. Thus, one way to analyze coercivity is through viscosity $\eta$ in our model \cite{Muller2016}. 
 
 In order to be as clear as possible, the details about the computation of the shear viscosity and entropy density ratio are present in Appendix \ref{visc}. In particular, we will focus on the $\eta/S$ ratio, where from Eq. \ref{viscosity} and Fig. \ref{p10}, we can analyze the dependence of the viscosity on the magnetic field, characterizing a magnetic side effect, and describing the slow relaxation of the magnetization of paramagnetic materials when they acquire magnetization in the presence of an external magnetic field $B$ (left panel of Fig. \ref{p10}). In the right panel,  we can observe that under an interval of the temperature $T$, the $\eta/S$ ratio is an increasing function when $B=0.$ 
\begin{figure}[!ht]
\begin{center}
\includegraphics[scale=0.71]{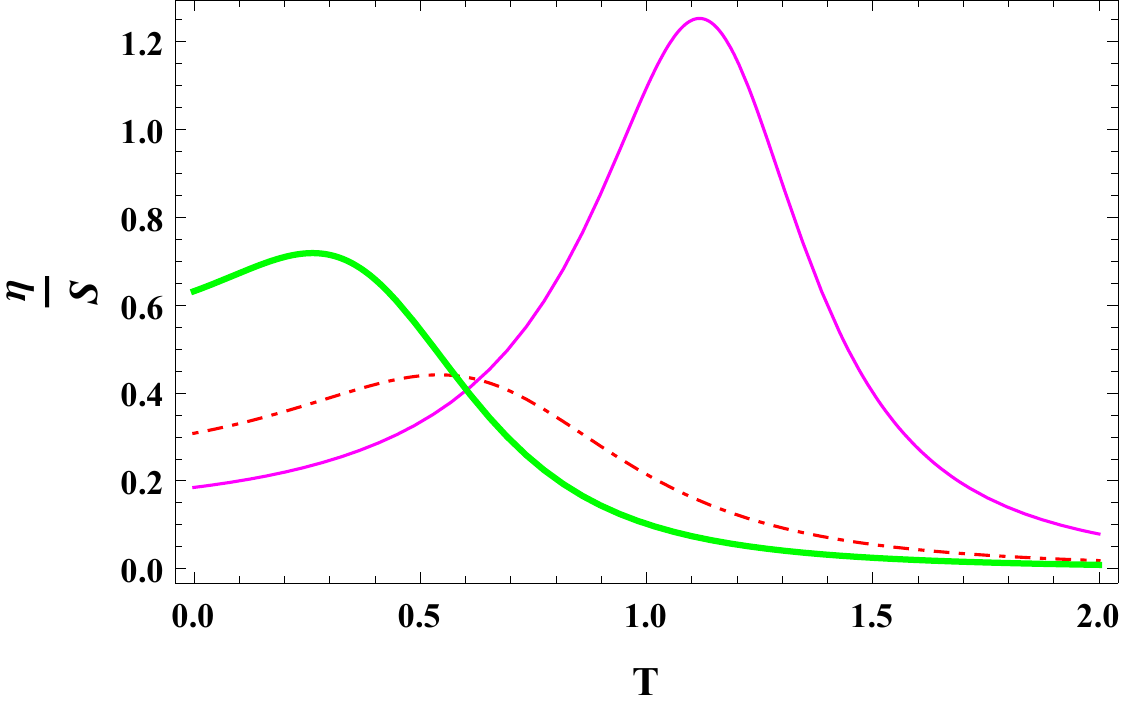}
\includegraphics[scale=0.71]{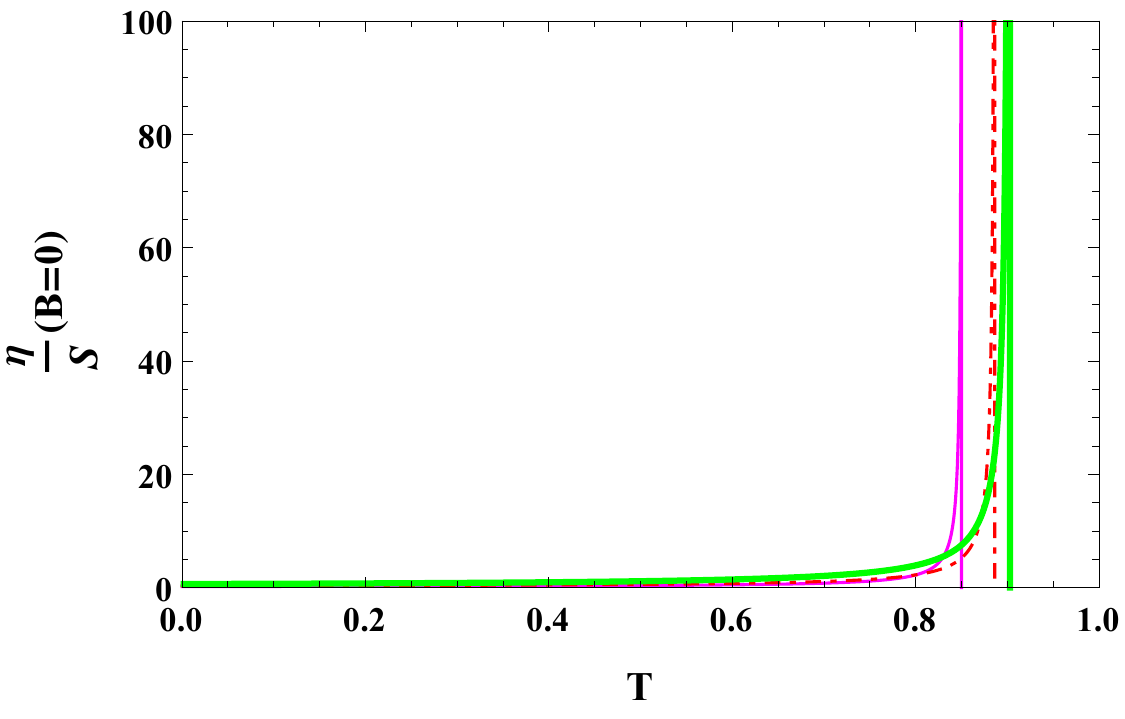}
\caption{{\sl Right panel:} The behavior of the $\eta/S$ ratio as a function of the temperature $T$ for different values for $\alpha=8/3$, $B=(4/5)T$, $\rho=1/4$, $\Lambda=-1$, $\gamma=1$ ({\sl pink curve}), $\gamma=2$ ({\sl red dot dashed curve}), $\gamma=2.5$ ({\sl green thick curve}). {\sl Left panel:} The behavior of $\eta/s$ for $B=0$.}\label{p10}
\end{center}
\end{figure}
}

\begin{figure}[!ht]
\begin{center}
\includegraphics[scale=0.71]{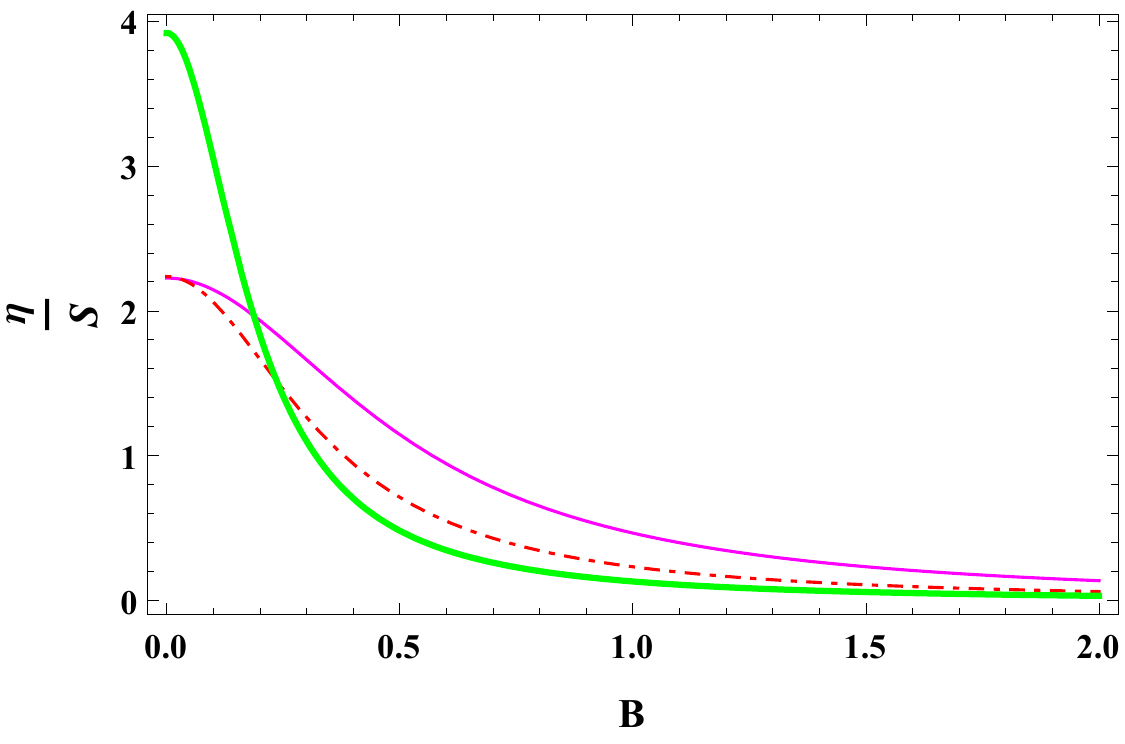}
\caption{The behavior of $\eta/S$ with respect to the magnetic field $B$, for different values for $\alpha=8/3$, $T=4/5$, $\rho=1/4$, $\Lambda=-1$, $\gamma=1$ ({\sl pink curve}), $\gamma=2$ ({\sl red dot dashed curve}), $\gamma=2.5$ ({\sl green thick curve}).}\label{p101}
\end{center}
\end{figure}

{On the other hand, and as we can see from  Fig. \ref{p101} at a temperature $T$ fixed when we observe as the paramagnetic material, represented by the RS brane, we can obtain a relation between $\eta/S$ with respect to the magnetic field $B$, which is a decreasing function. Here, when $B$ becomes large, we have that  $\eta/S\to 0$.}

We finalize this section showing the magnetic moment $N$ at a low temperature $T$, corresponding to order parameter $\rho$ in the absence of an external magnetic field, setting $B=0$, and then compute the value of $N$, defined as
\begin{eqnarray}
&&N=\dfrac{\lambda^{2}r_{h}}{2L}\int^{1}_{0}{\rho(r)dr}=-\frac{\lambda^{2}r_{h}}{2L}\left(-\dfrac{B}{m^{2}}+
\dfrac{1}{(\Delta_{+}+1)r^{\Delta_{+}}_{h}}+\dfrac{1}{(\Delta_{-}+1)r^{\Delta_{-}}_{h}}\right).\label{moment}
\end{eqnarray}

In Fig. \ref{p1}, it can be found that as the temperature decreases, the magnetization increases and the system is in the perfect order with the maximum magnetization at zero temperature. Thus, increasing the Horndeski parameters lowers the magnetization value and the critical temperature. Indeed, we have that the effect of a larger value of the parameters $\gamma$ and $m^{2}$ makes the magnetization harder and the ferromagnetic phase transition happen, which is in good agreement with previous works \cite{Zhang:2016nvj,Wu:2016uyj}. 

\begin{figure}[!ht]
\begin{center}
\includegraphics[scale=0.7]{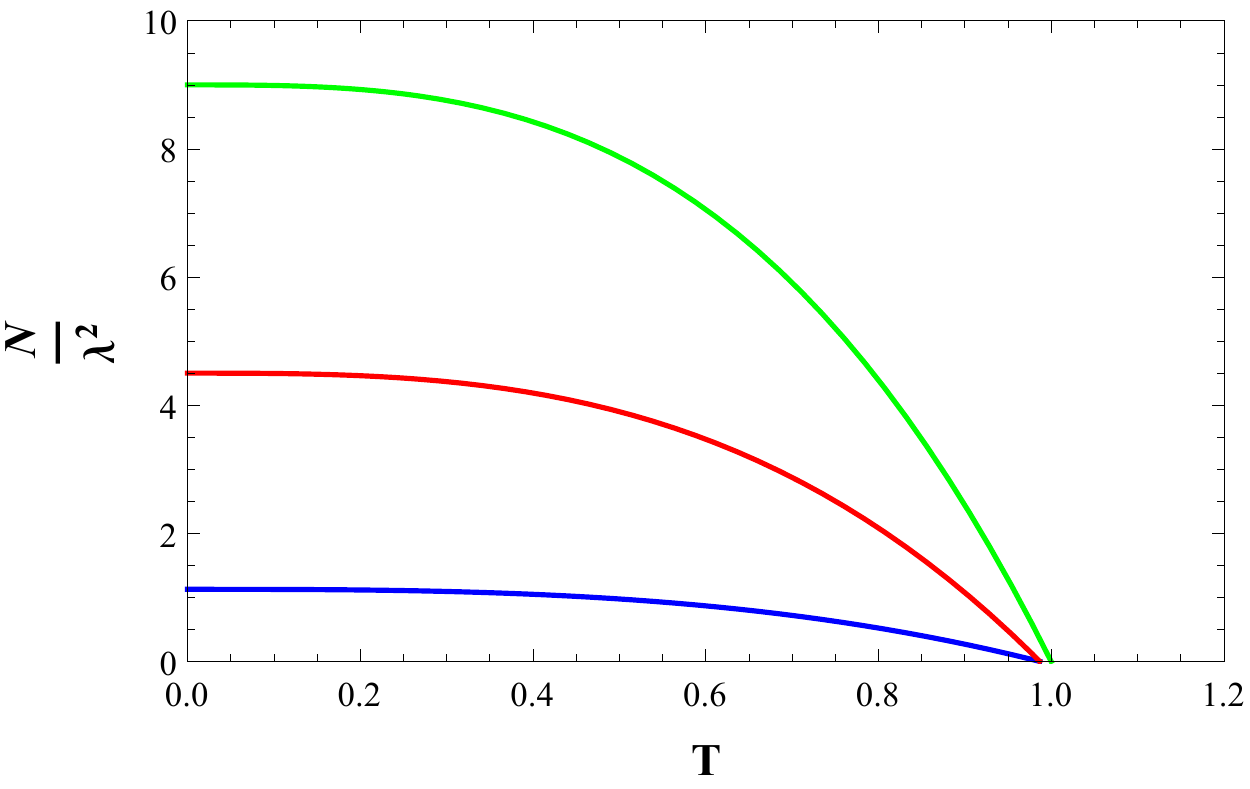}
\caption{The behavior of magnetic moment $N$ with different values for $B=0$, $\alpha=8/3$ with $\gamma=1;m^{2}=2$ ({\sl blue curve}), $\gamma=4;m^{2}=4$ ({\sl red curve}), $\gamma=8;m^{2}=6$ ({\sl green curve}). 
{\sl We consider in the Eq. \ref{moment} the transformations Eq.$\sim$(\ref{transfor}). }}\label{p1}
\end{center}
\end{figure}

Finally, we present the susceptibility density $\chi$ of the materials as a response to the magnetic moment. Thus, this behavior is an essential property of ferromagnetic materials. In order to study $\chi$ of the ferromagnetic materials in the Horndeski gravity and to consider the transformations Eq. (\ref{transfor}), we follow the definition

\begin{eqnarray}
&&\dfrac{\chi}{\lambda^{2}}=\lim_{B\to0}\dfrac{\partial N}{\partial B}=\left(\dfrac{3}{ 8 \pi m^{2}L^{2}}\right)\dfrac{1}{T}. \label{suscept}
\end{eqnarray}

\begin{figure}[!ht]
\begin{center}
\includegraphics[scale=0.6
]{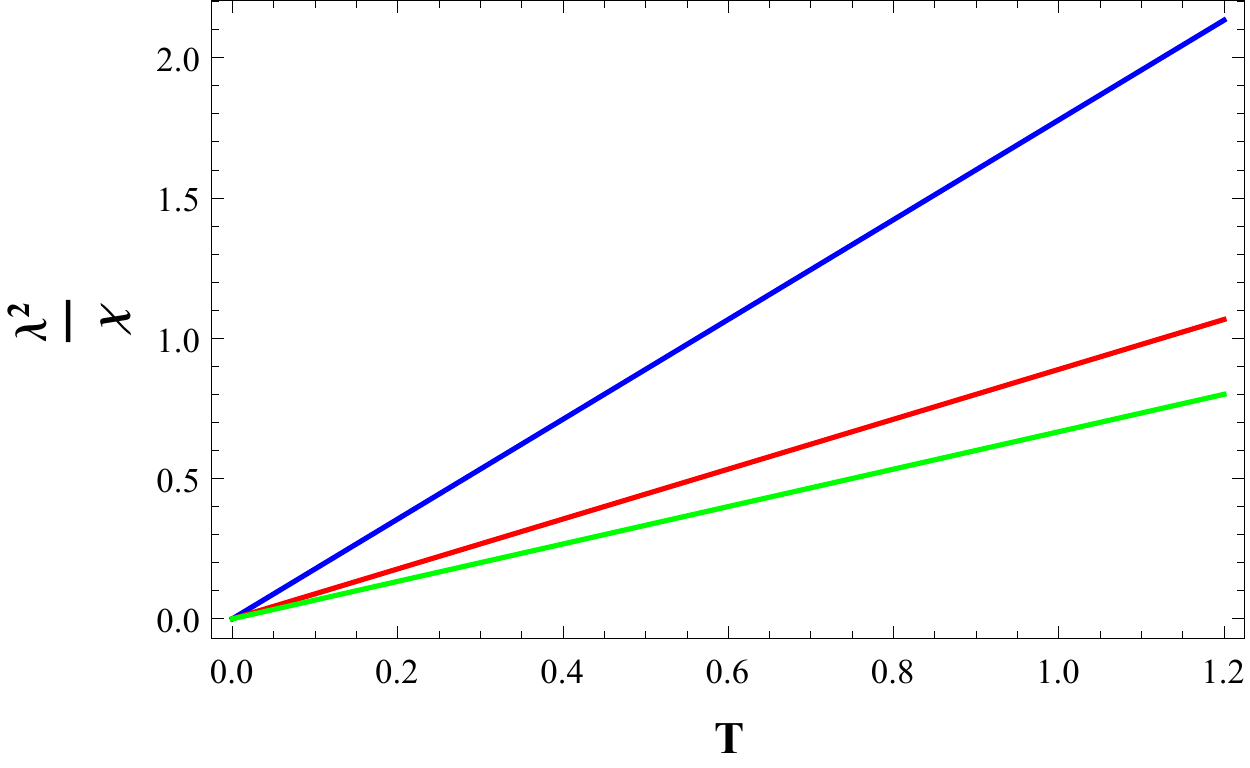}
\includegraphics[scale=0.6
]{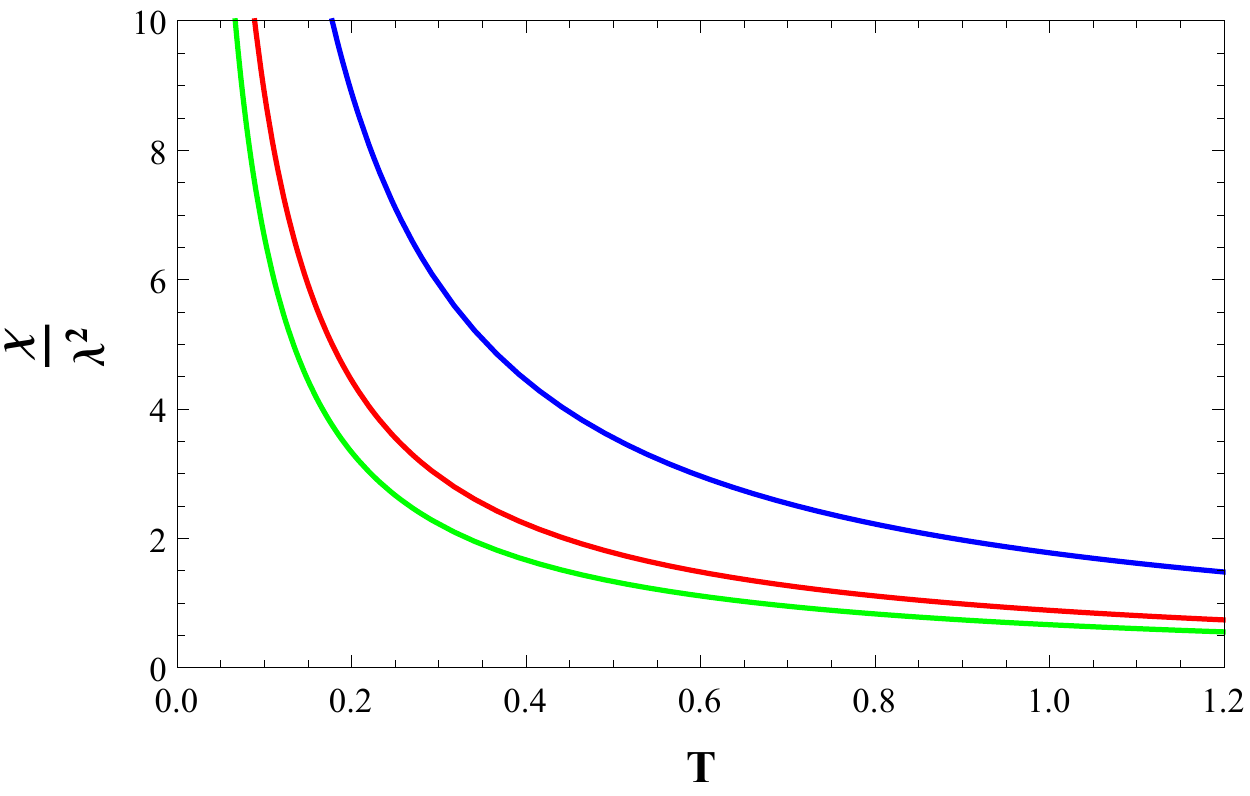}
\caption{The behavior of $1/\chi$ in the function of the temperature $T$ with different values for $\alpha=8/3$ with $\gamma=1;m^{2}=2$ ({\sl blue curve}), $\gamma=4;m^{2}=4$ ({\sl red curve}), $\gamma=8;m^{2}=6$ ({\sl green curve}). 
{\sl We consider in the Eq. (\ref{suscept}) the transformations given in Eq.(\ref{transfor}).}}\label{p2}
\end{center}
\end{figure}

In Fig. \ref{p2}, we have the behavior of $1/\chi$ and $\chi$ as a function of the temperature $T$ for different choices of $m^2$ and $\gamma$. In our case, in the {\sl right panel}, we have that increasing each one of these parameters makes the susceptibility value decrease when the temperature increases. This fact agrees with our expectation of paramagnetic materials because when we remove the external magnetic field, the paramagnetic substance loses its magnetism. Its magnetic susceptibility is very small, but positive, and decreases with increasing temperature. In fact, this magnetic susceptibility is only part of the background black hole and the other part of the polarization field. For pure dyonic Reissner-Nordström-AdS black hole, we have a diamagnetic material. In this sense, in the chemical reference, we have that a particle (atom, ion, or molecule) is paramagnetic or diamagnetic when the electrons in the particle are paired due to the external magnetic field \cite{Zhang:2016nvj,Wu:2016uyj}.

\section{Conclusions and discussions}\label{v4}

In four dimensions, we analyzed an AdS/BCFT model of a condensed matter system at finite temperature and charge density living on a 2+1-dimensional space with a boundary, showing an extension of the previous work presented in \cite{Santos:2021orr}, where in addition to the contributions of the theory together with the boundary terms, we include the components $A_{\mu}$ and $M_{\mu\nu}$, responsible to construct the ferromagnetic/paramagnetic model.

Via the resolution of the field equations, and using the no-hair theorem, we extend to the four-dimensional configuration obtained in
\cite{Santos:2021orr,Bravo-Gaete:2014haa}. From the above solution, we present the $Q$ profile, found a numerical solution, and present it in Fig. \ref{p1}, where the Horndeski parameter $\gamma$ takes an important role. Together with the above, we show that components of  $M_{\mu\nu}$ can be viewed as dual fields of the order parameter in the paraelectric/ferroelectric phase transition in dielectric materials. Through the NBC over $n^{\mu}M|_{Q}$,  we found  the ratio $\rho/B$, where for some particular cases is a constant proportional to a ratio of the coefficients appearing in the gravity action. These properties resemble a quantum Hall system, which suggests at the boundary $Q$ in the ($\rho, B$) plane will be a localized condensate. 

Additionally, via the solution we performed a holographic renormalization, calculating the Euclidean on-shell action, which is related to the free energy $\Omega$, and allowing us to obtain the entropy $S$ and the heat capacities $C_V, C_P$, thanks to the contribution to the bulk as well as the boundary. With respect to the entropy $S$, we show that when the magnetic field is present we see it exhibits similar behavior as for example ferromagnetic materials with nearly zero coercivity and hysteresis. Nevertheless,  when $B=0$ the disorder entropy of the magnetic moments increases, being a characteristic of paramagnetism. Together with the above, with respect to $C_V$ and $C_P$, we obtained for both cases stable and unstable 
phases, due to the spontaneous electric polarization, which was realized in our model from the application of the
magnetic external field $B$, being influence via the Horndeski gravity, represented through $\gamma$. We also show that the specific heat $C_P$ behaves like a material of the type $DyAl_{2}$, having a growth behavior similar to that expected from the experimental point of view, as presented by \cite{PhysRevB.72.024403}.


Currently, we can observe that the microscopic differences between real experimental systems, in relation to theories with gravitational dual suggest that, in the near future, we will have measurements of these values for experimental quantities obtained holographically. So many measurements can realistically aspire to more than useful benchmarks. Furthermore, it is important to highlight in this regard the need to take the big limit $N$ in holographic calculations \cite{Maldacena:1997re}. We now have a clarity of the value of the ratio between shear viscosity and entropy density, $\eta/S=1/4\pi$, which is universal in classical gravity to usual classical gravity \cite{Kovtun:2004de}. Furthermore, in the Horndeski gravity, these relations are modified by the parameter $\gamma$. However, there are controlled corrections $1/N$ for this result, which can be both positive and negative and which for realistic values of $N$ show significant changes in the numerical value of the ratio. As we show in our model, the violation of this universal bound in the Horndeski gravity with gauge fields changes the $\eta/S$ ratio (see Fig.\ref{p10} and Fig.\ref{p101}), where this behavior is similar to the results of \cite{Hartnoll:2016tri,Alberte:2016xja,Baggioli:2017ojd,Baggioli:2020ljz}. Furthermore, as $\gamma$ increases, we can observe a translational symmetry breaking that survives the lower energy scales. According to Fig. \ref{p101}, we have $\eta/S\to 0$ at low temperatures.

One of the strongest motivations for working with AdS/BCFT for condensed matter physics rests on two pillars. The first is that, although theories with holographic duals may exhibit specific exotic features, they also have features that are expected to be generic to tightly coupled theories, for example, the quantum critiques. In this sense, theories with gravitational duals are computationally tractable examples of generic tightly coupled field theories, and we can use them both to test our generic expectations and to guide us in refining those expectations. Thus, the examples discussed here are special cases of the fact that real-time finite temperature transport is much easier to calculate via AdS/BCFT than almost any other microscopic theory.

\acknowledgments
F.S. would like to thank the group of Instituto de Física da UFRJ for fruitful discussions about the paramagnetic systems. In special to the E. Capossoli, Diego M. Rodrigues and Henrique Boschi-Filho. S.O. performed the work in the frame of the "Mathematical modeling in interdisciplinary research of processes and systems based on intelligent supercomputer, grid and cloud technologies" program of the NAS of Ukraine.  M.B. is supported by PROYECTO INTERNO UCM-IN-22204, L\'iNEA REGULAR. 


\begin{appendix}

\section{Shear viscosity and entropy density ratio with magnetic field}\label{visc}
We will present the calculation of the ratio $\eta/S$ following the procedures  \cite{Brito:2019ose,Feng:2015oea,Bravo-Gaete:2022lno,Kovtun:2004de,Hartnoll:2016tri}. For this, we consider a perturbation along the $xy$ direction in the metric Eq.\ref{ansatz} \cite{Brito:2019ose,Feng:2015oea}, in this sense, we have
\begin{eqnarray}
ds^{2}=\frac{L^{2}}{r^{2}}\left(-f(r)dt^{2}+dx^{2}+dy^{2}+2\Psi(r,t)dxdy+\frac{dr^{2}}{f(r)}\right).\label{pertur}
\end{eqnarray}
From the overview point of the holographic dictionary, this procedure maps the fluctuation of the diagonal in the bulk metric in the off-diagonal components of the dual energy-momentum tensor. In this sense, we have a linear regime where fluctuations are associated with shear waves in the boundary fluid. Thus, substituting this metric (\ref{pertur}) in the Horndeski equation (${\cal E}_{\mu\nu}=0$) for $\mu=x$ and $\nu=y$,  one obtains:
\begin{eqnarray}
&&{\cal P}_{1}\Psi^{''}(r,t)+{\cal P}_{2}\Psi^{'}(r,t)+{\cal P}_{3}\ddot{\Psi}(r,t)=0\,,
\end{eqnarray}
where we defined
\begin{eqnarray}
&&{\cal P}_{1}=9\gamma^{2}(\alpha-\gamma\Lambda)f^{2}(r),\qquad {\cal P}_{2}=-3\gamma(\alpha-\gamma\Lambda)f(r)(2\alpha L^{2}-6\gamma r^{3}/r^{3}_{h}),\cr
&&{\cal P}_{3}=-9\gamma^{2}r(3\alpha+\gamma\Lambda).
\end{eqnarray}
Using the ansatz: 
\begin{eqnarray}
&&\Psi(r,t)=e^{-i\omega t}\Phi(r),\\
&&\Phi(r)=\exp\left(-i\omega K\ln\left(\frac{6\gamma^{2}r^{3}f(r)}{{\cal G}}\right)\right),\quad\,{\cal G}=\frac{L^{2}V}{G_{N}}\left(1-\frac{\xi}{4}\right),
\end{eqnarray}
we obtain
\begin{eqnarray}
K=\frac{1}{4\pi T}\sqrt{\frac{3\alpha+\gamma\Lambda}{\alpha-\gamma\Lambda}},
\end{eqnarray}
with $T$ the Hawking temperature given previously in (\ref{eq:Th}). At this point, we must evaluate the Lagrangian (\ref{eq:Lhorn}), using the metric function from (\ref{L16}), and expand it up to quadratic terms in $\Psi$ and its derivatives  \cite{Feng:2015oea}. In this way, we can study the boundary field theory using the AdS/CFT correspondence where the quadratic terms in the Lagrangian, after removing the second derivative contributions using the Gibbons-Hawking term, can be written as

\begin{eqnarray}
&&{\cal H}_{shear}=P_{1}\Psi^{2}(r,t)+P_{2}\dot{\Psi}(r,t)+P_{3}\Psi^{'2}(r,t)+P_{4}\Psi(r,t)\Psi^{'}(r,t),
\end{eqnarray}
where
\begin{eqnarray}
&&P_{1}=-\frac{48L^{2}}{9r^{7}f(r)},\qquad P_{2}=\frac{4\gamma\,L^{2}}{r^{7}},\qquad P_{3}=\frac{6\gamma^{2}}{r^{3}f(r)},\qquad
P_{4}=(\alpha+\gamma\Lambda)\frac{2\gamma^{2}L^{4}}{\alpha\,r^{7}f(r)}.
\end{eqnarray}
Here, $ (\dot{\phantom{a}} )$ denotes the derivative with respect $t$. Finally, viscosity $\eta$ is determined from the  term $P_{3}\Psi(r,t)\Psi^{'}(r,t)$ which reads
\begin{eqnarray}
\eta=\frac{1}{4\pi}\frac{{\cal G}}{4r^{2}_{h}}\sqrt{\frac{3\alpha+\gamma\Lambda}{\alpha-\gamma\Lambda}},
\end{eqnarray}
where the entropy, from (\ref{eq:ent-total})-(\ref{BT8}), can be written as
\begin{eqnarray}
S=\frac{{\cal G}\mathcal{F}}{4r^{2}_{h}},
\end{eqnarray}
with
\begin{eqnarray*}
\mathcal{F}&=&1+\left(\frac{B^{2}\cos^{2}(\theta{'})b(\theta{'})}{5m^{2}\rho^{2}}\left(\frac{4\pi T}{3}\right)^{4}+\frac{q(\theta^{'})}{4}\left(\frac{4\pi T}{3}\right)^{2}\right)\nonumber\\
&-&\frac{\sec(\theta{'})}{\left(1-\frac{\xi}{4}\right)}\left(-\frac{B^{2}\cos^{2}(\theta{'})b(\theta{'})}{2m^{2}\rho^{2}}\left(\frac{4\pi T}{3}\right)^{3}+\frac{q(\theta^{'})}{2}\left(\frac{4\pi T}{3}\right)\right),
\end{eqnarray*}
and $T$ given in (\ref{eq:Th}). Thus, after algebraic manipulation and imposing $V=1$, we have:
\begin{eqnarray}
&&\frac{\eta}{S}=\frac{1}{4\pi\mathcal{F}}\sqrt{\frac{3\alpha+\gamma\Lambda}{\alpha-\gamma\Lambda}},\label{viscosity}
\end{eqnarray}
where $B=0$ and $\theta{'}=\pi/2$, we recover the result of \cite{Feng:2015oea}.

\end{appendix}


\end{document}